\documentclass[11pt]{article}

\pdfoutput=1

\usepackage{amsmath,amssymb,amsfonts,amscd,mathrsfs,colonequals,mathtools}

\usepackage[utf8]{inputenc}

\usepackage[]{graphicx}
\usepackage{dsfont} 
\usepackage{verbatim}

\usepackage{xcolor}
\definecolor{darkblue}{rgb}{0.1,0.1,.7}
\usepackage[colorlinks, linkcolor=darkblue, citecolor=darkblue, urlcolor=darkblue, linktocpage]{hyperref} 
\usepackage[square, comma, compress,numbers]{natbib}
\usepackage{geometry}
\usepackage[margin=10pt,font=small,labelfont=bf]{caption}

\def\th{\tfrac{1}{2}}
\def\Ksplit{K_\text{split}}

\def\ee{\end{equation}}

\def\dps{\displaystyle}
\def\la{\lambda}
\def\msf{\mathsf}

\def\wh{\widehat}

\def\wt{\widetilde}
\def\id{\text{id}}
\def\rd{{d}}
\newcommand{\hf}{\frac{1}{2}}
\def\bsub{\begin{subequations}}
\def\esub{\end{subequations}}

\def\lra{\leftrightarrow}
\def\sl2{SL(2,\mathbb{R})} 
\def\Fphi{F_{\phi \phi \phi \phi}}
\def\Fsch{F_{\phi_1 \phi_2 \phi_3 \phi_4}}
\def\Ftch{F_{\phi_3 \phi_2 \phi_1 \phi_4}}
\def\wils{\mfr{p}}

\newcommand{\ud}[2]{^{#1}_{\phantom{#1}#2}}

\def\dd{\delta}

\newcommand{\floor}[1]{\left \lfloor #1 \right \rfloor }

\newcommand{\expec}[1]{\langle #1 \rangle}
\newcommand{\bexpec}[1]{\big\langle #1 \big\rangle}
\def\ldef{\mathrel{\mathop:}=}

\newcommand{\reef}[1]{(\ref{#1})}
\newcommand{\redeqn}{equation}
\def\beq{\begin{\redeqn}} 
\def\eeq{\end{\redeqn}}
\def\nn{\nonumber} 

\def\mbb{\mathbb}

\def\mca{\mathcal}
\def\mfr{\mathfrak}
\def\mrm{\mathrm}
\def\msc{\mathscr}
\def\qaq{\quad \text{and} \quad}

\def\pd{\partial}
\def\a{\alpha}
\def\b{\beta}

\def\la{\lambda}

\def\G{\Gamma}

\def\Oo{\mathcal{O}}

\def\eps{\epsilon}

\def\unit{\mathds{1}} 

\newcommand{\fracpt}[2]{\frac{\partial #1}{\partial #2}}

\newcommand{\limu}[1]{\mathrel{\mathop{\sim}\limits_{\scriptstyle{#1}}}}

\def\half{\frac{1}{2}}
\def\thalf{\tfrac{1}{2}}

\numberwithin{equation}{section}

\vspace{-2cm}

\begin{document}

\vspace*{-.6in} \thispagestyle{empty}
\begin{flushright}
YITP-SB-17-7
\end{flushright}
\vspace{1cm} {\Large
\begin{center}
  {\bf Crossing Symmetry in Alpha Space}
\end{center}}
\vspace{1cm}
\begin{center}
{\bf Matthijs Hogervorst$^{a}$, Balt C.~van Rees$^{b}$ }\\[2cm] 
{
$^{a}$ C.N.\@ Yang Institute for Theoretical Physics, Stony Brook University, USA\\
$^{b}$ Centre for Particle Theory \& Department of Mathematical Sciences,\\
Durham University, Durham, UK
}
\\
\end{center}
\vspace{14mm}

\begin{abstract}
  We initiate the study of the conformal bootstrap using Sturm-Liouville theory, specializing to four-point functions in one-dimensional CFTs. We do so by decomposing conformal correlators using a basis of eigenfunctions of the Casimir which are labeled by a complex number $\alpha$.  This leads to a systematic method for computing conformal block decompositions. Analyzing bootstrap equations in alpha space turns crossing symmetry into an eigenvalue problem for an integral operator $\msf{K}$. The operator $\msf{K}$ is closely related to the Wilson transform, and some of its eigenfunctions can be found in closed form.
  
\end{abstract}
\vspace{12mm}

\newpage

{
\setlength{\parskip}{0.05in}
\tableofcontents
}

\setlength{\parskip}{0.05in}

\section{Introduction}

Symmetries and consistency conditions play an important role in quantum field theory. This is especially true in the realm of Conformal Field Theories (CFTs), which can be analyzed by combining constraints from  conformal invariance, unitarity and crossing symmetry. This set of ideas is known as the conformal bootstrap~\cite{Ferrara:1973vz,Ferrara:1973yt,Polyakov:1974gs,Belavin:1984vu}. It was revived in~\cite{Rattazzi:2008pe} and has led to a wealth of numerical and analytical results about CFTs, see for instance~\cite{Heemskerk:2009pn,Rattazzi:2010yc,Poland:2011ey,ElShowk:2012ht,Pappadopulo:2012jk,ElShowk:2012hu,Fitzpatrick:2012yx,Komargodski:2012ek,Gliozzi:2013ysa,Beem:2013sza,El-Showk:2014dwa,Kos:2014bka,Alday:2015ota,Hartman:2015lfa,Kim:2015oca,Chester:2016wrc,Dey:2016zbg,Hofman:2016awc,Kos:2016ysd,El-Showk:2016mxr,Alday:2016njk,Mazac:2016qev,Simmons-Duffin:2016wlq}.\footnote{See~\cite{Rychkov:2016iqz,Qualls:2015qjb,Simmons-Duffin:2016gjk} for an introductory discussion of the conformal bootstrap.} 

Since the bootstrap uses constraints coming from correlation functions, it is natural to express crossing symmetry as a sum rule in position space. This is not strictly necessary: for instance, some properties of CFT correlators are more transparent in Mellin space~\cite{mack1,Penedones:2010ue,Fitzpatrick:2011ia,Fitzpatrick:2011hu,Fitzpatrick:2011dm,Costa:2012cb,Goncalves:2014rfa,Gopakumar:2016wkt,Gopakumar:2016cpb}. In the present paper we introduce \emph{alpha space}, an integral transform for CFT correlators based on the Sturm-Liouville theory of the conformal Casimir operator. As we will explain, alpha space can be used to rephrase crossing symmetry as an eigenvalue problem.

To illustrate this idea, consider the toy crossing equation
\beq
\label{eq:cross0}
\sum_{n=0}^5 c_n \, p_n(z) = \sum_{n=0}^5 c_n \, p_n(1-z)
\eeq
involving the following polynomials:\footnote{Up to a choice of normalization, these are the Kravchuk polynomials with $N=5$ and $p = 1/2$~\cite{AAR,askeyScheme}.}
\beq
p_n(z) = (-1)^n \sum_{j=0}^n 2^j \binom{5-z}{n-j} \binom{z}{j}\,,
\qquad
n = 0,\ldots,5\,.
\eeq
How can we determine the set of all $c_n$ that satisfy~\reef{eq:cross0}?  Since the $p_n(z)$ are polynomials, various brute-force methods can be used. More elegantly, we can realize that the $p_n$ form a complete basis for the space of polynomials of degree $\leq 5$, orthogonal with respect to the inner product
\beq
\label{eq:ipk}
\int\!d \mu  f(z) g(z)\,,
\qquad
\int\!d \mu = \sum_{k=0}^5 \frac{(-1)^k}{2^k} \binom{5}{k}  \int\!dz \, \delta(z-k)\,.
\eeq
This implies that the $p_n(1-z)$ appearing in the RHS of the crossing equation can be decomposed as follows:
\beq
p_n(1-z) = \sum_{m=0}^5 \msf{Q}\ud{m}{n} \, p_{m}(z)
\eeq
for some $6 \times 6$ matrix $\msf{Q}$. The latter can be easily computed using~\reef{eq:ipk}. Since $z \mapsto 1-z$ is an involution, we must have $\msf{Q}^2 = \unit_{6 \times 6}$, as can be checked easily.  Eq.~\reef{eq:cross0} can now be recast as
\beq
\label{eq:finmat}
c_n = (\msf{Q} \cdot c)_n 
\eeq
hence our problem reduces to finding all eigenvectors of $\msf{Q}$ with eigenvalue $+1$. There are three such eigenvectors:
\beq
\label{eq:ff1}
f_1 = p_0 \, (\equiv 1)
\quad
f_2 = 2p_1 - p_3 - p_4\,,
\quad
f_3 = p_2 +2p_3 + 2 p_4\,,
\eeq
so the most general solution to~\reef{eq:cross0} is
\beq
\sum_{n=0}^5 c_n p_n(z) = \sum_{i=1}^3 t_i f_i(z), \quad t_i \in \mbb{R}\,.
\eeq

In this paper we consider one-dimensional (defect) CFTs which are governed by crossing equations similar to~\reef{eq:cross0}. For definiteness, let us consider a four-point function $F(z)$ of identical operators of dimension $h_\phi$, admitting a conformal block decomposition
\beq
\label{eq:Fdef}
F(z) = \int_{0}^\infty\!d h\,  \rho(h) k_h(z)\,,
\quad
\rho(h) = \sum_n c_n \delta(h-h_n)\,,
\eeq
where the $k_h(z)$ are $\sl2$ conformal blocks:
\beq
\label{eq:CB}
k_h(z) = z^h {}_2F_1(h,h;2h;z)\,.
\eeq
The spectrum $\{h_n\}$ and the OPE coefficients $c_n \geq 0$ are typically unknown. Bootstrapping entails computing or constraining these CFT data using the crossing relation
\beq
\label{eq:cross1}
F(z) = \left(\frac{z}{1-z}\right)^{2h_\phi} F(1-z)\,.
\eeq
There are various technical differences between this $d=1$ bootstrap problem and the previous toy example. For one, $h$ takes its values in the continuum $\mbb{R}_{\geq 0}$, whereas the toy example had a finite and discrete spectrum. Nevertheless, it is tantalizing to apply the logic from the toy example to the bootstrap. For instance, one could hope to constrain the density $\rho(h)$ from~\reef{eq:Fdef} through a relation of the form
\beq
\label{eq:wrong}
\rho(h) \stackrel{?}{=} \int_0^\infty\!d {h}' \, \msc{Q}(h,{h}'|h_\phi) \rho({h}')
\eeq
for some continuous kernel $\msc{Q}(h,{h}'|h_\phi)$ which plays the role of $\msf{Q}$. Sadly Eq.~\reef{eq:wrong} cannot quite be true. The reason is that the conformal blocks $k_h(z)$ don't form an orthogonal basis of functions on $(0,1)$. The principal aim of this paper is to demonstrate that it is nevertheless possible to write down a qualitatively very similar relation. In order to do so we use a new basis of functions to transform our four-point function to a space that we denote as {alpha space}. In this space we can properly define \eqref{eq:wrong} in terms of a \emph{crossing symmetry kernel} $K$ which we will explicitly compute. We will discuss its main features and explain how the ordinary conformal block decomposition is recovered from an analytic continuation in alpha.

We stress that the philosophy of studying CFTs using crossing kernels --- \emph{\`{a} la}~\reef{eq:wrong} --- is  not new. An early avatar  of this idea can be found in Eq.~(2.66) of Ref.~\cite{Dobrev:1975ru}. Nonetheless, we are not aware of earlier work where the relevant $SO(d,2)$ or $\sl2$ crossing kernels have been worked out in detail. An exception is the 2$d$ Liouville CFT, for which the crossing kernels have been computed~\cite{Ponsot:1999uf,Ponsot:2000mt} as the $6-j$ symbol of a class of representations of $U_q(\mathfrak{sl}(2,\mbb{R}))$, leading to a formal proof of consistency of the theory.\footnote{See also~\cite{Hadasz:2013bwa,Pawelkiewicz:2013wga,Esterlis:2016psv}.} The case of {rational} 2$d$ CFTs (i.e.\@ Virasoro minimal models) is also of interest, since in such theories the crossing kernel is realized as a finite matrix~\cite{Moore:1988qv,Moore:1989vd,AlvarezGaume:1989vk}.  We will comment on the group-theoretic interpretation of our crossing symmetry kernel in Sec.~\ref{sec:discussion}.

The outline of this paper is as follows. In Sec.~\ref{sec:1d} we review the one-dimensional bootstrap problem and solve the Sturm-Liouville problem for the $\sl2$ Casimir operator. This allows us to construct a complete, orthogonal basis of eigenfunctions on the interval $(0,1)$. In Sec.~\ref{sec:kernel} we use these basis functions to derive a crossing equation similar to~\reef{eq:wrong}, and we study the properties of the relevant kernel $K$. Sec.~\ref{sec:appli} describes several possible applications of crossing kernels to the conformal bootstrap.

{\it Note added:} while preparing this manuscript we learned about Ref.~\cite{gadde}, which  discusses a crossing kernel approach to both $SU(2)$ and conformal crossing symmetry equations and is tangentially related to this paper.

\section{One-dimensional bootstrap and alpha space}\label{sec:1d}

This section is devoted to the Sturm-Liouville theory of the conformal Casimir of $\sl2$, the conformal group in one spacetime dimension. One-dimensional CFTs arise in the description of line defects in higher-dimensional theories~\cite{Gaiotto:2013nva,Billo:2013jda,Gliozzi:2015qsa,Billo:2016cpy}. Although 1$d$ CFTs are in many ways more tractable than $d$-dimensional systems, we also note that many salient features of the $d$-dimensional bootstrap already appear at the level of $d=1$. In addition the 1$d$ conformal blocks appear naturally in the light-cone limit of the higher-dimensional crossing symmetry equations, where it becomes possible to obtain non-trivial analytic results~\cite{Alday:2007mf,Fitzpatrick:2012yx,Komargodski:2012ek}.

\subsection{Sturm-Liouville theory of the $\sl2$ Casimir}
\label{sec:SL2SL}

We will start by analyzing the four-point function of a single primary (or lowest-weight) operator $\phi(x)$ in a 1$d$ CFT. The general case will be addressed in Sec.~\ref{sec:mixed}.  The only quantum number of $\phi$ is its scaling dimension $h_\phi$, and conformal symmetry dictates that $\expec{\phi \phi \phi \phi}$ has the following form:
\beq
\expec{\phi(x_1) \phi(x_2) \phi(x_3) \phi(x_4)} = \frac{\Fphi(z)}{|x_1 - x_2|^{2h_\phi}  |x_3 - x_4|^{2h_\phi}}
\eeq
where the points $x_i \in \mbb{R}$ lie on a line and $z$ is the following cross ratio:
\beq
\label{eq:zdef}
z \ldef \frac{|x_{12}| |x_{34}|}{|x_{13}| |x_{24}|} \in (0,1)
\eeq
writing $x_{ij} = x_i - x_j$.\footnote{Although {\it a priori} the variable $z$ is not restricted to the unit interval, we require $z \in (0,1)$ to guarantee OPE convergence on both sides of the bootstrap equation.} The function $\Fphi(z)$ admits the following conformal block (CB) decomposition:
\beq
\label{eq:CBdec}
\Fphi(z) = \sum_\Oo \, \la_{\phi \phi \Oo}^2 \, k_{h_\Oo}(z)
\eeq
where the functions $k_{h}(z)$ are the 1$d$ conformal blocks defined in Eq.~\reef{eq:CB}. The sum runs over all operators $\Oo$ in the $\phi \times \phi$ OPE of dimension $h_\Oo$, and $\la_{\phi \phi \Oo}$ is the $\Oo \in \phi \times \phi$ OPE coefficient.  
Finally, crossing symmetry (invariance under the exchange $x_i \lra x_j$) of the $\expec{\phi \phi \phi \phi}$ correlator leads to the bootstrap constraint
\beq
\label{eq:1dBS}
\Fphi(z) = \left(\frac{z}{1-z} \right)^{2h_\phi} \Fphi(1-z)
\eeq
which must hold for all $0 \leq z \leq 1$. 

We will not assume unitarity (i.e.\@ reflection positivity) in this paper. Just for completeness, we recall that if the CFT in question is unitary, the decomposition~\reef{eq:CBdec} is constrained as follows:
\begin{itemize}
  \item the $\la_{\phi \phi \Oo}$ must be real-valued, hence $\la_{\phi \phi \Oo}^2 > 0$;
  \item there must be a contribution of the unit operator $\unit$ with $h_\unit = 0$ and $\la_{\phi \phi \unit} = 1$;
  \item all other operators (including $\phi$) have $h_\Oo > 0$.
\end{itemize}

As noted in the introduction, it is conventional in the CFT literature to investigate the bootstrap equation~\reef{eq:1dBS} in position space. Here we will take a different approach. We start by remarking that the conformal blocks $k_h(z)$ are eigenfunctions of a second-order differential operator $D$, the quadratic Casimir operator of $\sl2$:
\beq
\label{eq:sl2}
D \cdot k_h(z) = h(h-1) k_h(z)\,,
\qquad
D  =  z^2(1-z) \pd^2 - z^2 \pd\,. \\
\eeq
In what follows, we will develop the Sturm-Liouville theory of the operator $D$ on the interval $(0,1)$.\footnote{{\bf Note added}: although we have not attempted to do so, it is in principle possible to change the boundary conditions at $z=1$ \cite{littlejohn2011legendre}. We thank Miguel Paulos for pointing out this reference.}  As a first step, we notice that $D$ can be written in the following suggestive form:
\beq
\label{eq:Drecast}
D \cdot f(z) = z^2 \fracpt{}{z} \left[(1-z) f'(z) \right].
\eeq
This implies that $D$ is self-adjoint with respect to the inner product
\beq
\label{eq:innerProd}
\bexpec{ f,g }  = \int_0^1  \frac{d z}{z^2} \; \overline{f(z)} g(z)
\eeq
where $f,g$ are functions $(0,1) \to \mbb{C}$ that are well-behaved near $z=0$ and $z=1$. Indeed, we have
\beq
\bexpec{ f, D \cdot g } - \bexpec{ D\cdot f,  g }  = \int_0^1 \!d z  \, \fracpt{}{z} \left[ (1-z) (\overline{f}g' - \overline{f'}g) \right]
\eeq
which is a boundary term. Of course, not all functions have a finite norm with respect to the inner product~\reef{eq:innerProd}. Requiring that a function $f$ is square integrable leads to the following constraints on its asymptotics near $z=0$ and $z=1$:
\beq
\label{eq:L2asymp}
f(z) \; \limu{z \to 0} \; z^{1/2+ \eps}
\quad
\text{and}
\quad
f(z) \; \limu{z \to 1} \; (1-z)^{-1/2+\eps'}
\eeq
for constants $\eps, \eps' > 0$. In particular, this implies that in a unitary CFT all four-point functions $\Fphi(z)$ have a divergent norm with respect to~\reef{eq:innerProd}.

\begin{figure}[htbp]
  \begin{center}
    \includegraphics[scale=.65]{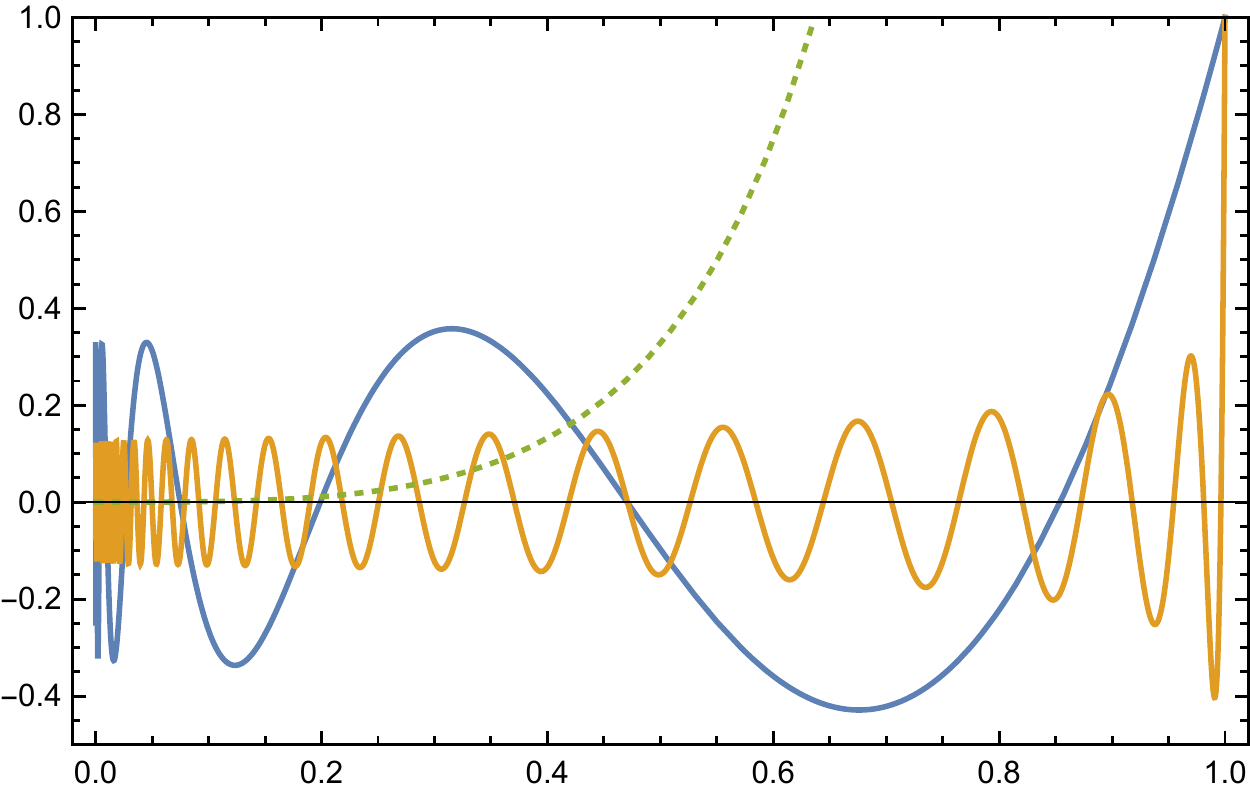}
   \caption{Plot of $z^{-1/2} \, \Psi_\a(z)$ for $\a = 2i$ (blue) and $\a = 20i$ (orange), as well as $k_h(z)$ for $h=3$ (dotted green). Both the oscillatory behaviour of the $\Psi_\a(z)$ near $z=0$ and their $O(\sqrt{z})$ growth are clearly visible. \label{fig:plotfuncs}}
\end{center}
\end{figure}

Our next order of business is to construct an orthogonal basis of eigenfunctions of $D$. We start by solving the eigenvalue equation $D \cdot f = \la f$. After writing $\la = \a^2 - 1/4$ for convenience, we find that the general solution (for $\a \neq 0$) is given by
\beq
\label{eq:ans1}
f(z) = A_1(\a) k_{\half+\a}(z) + A_2(\a)k_{\half-\a}(z)
\eeq
for two constants $A_{1,2}(\a)$ that are to be determined. 
In order to fix them, let's analyze the $z \to 0,1$ asymptotics of $f(z)$. First, we notice that the blocks themselves are logarithmically divergent near $z = 1$. To be precise, we have
\beq
k_{\half+\a}(z) \; \limu{z \to 1} \; - \frac{\Gamma(1+2\a)}{\Gamma^2(\thalf+\a)} \ln(1-z) \, + \, \text{regular}
\eeq
and likewise for $k_{\half-\a}(z)$. 
Requiring that~\reef{eq:ans1} has a finite limit as $z \to 1$ therefore determines the relative coefficient $A_1(\a)/A_2(\a)$. Fixing the overall normalization by imposing $f(1) = 1$, we arrive at the following eigenfunctions:\footnote{Remarkably, these are \emph{not} the usual `shadow-symmetric' blocks obtained by integrating one-dimensional three-point functions over the real axis \cite{Dolan:2000ut}. Indeed, in one dimension this integral is easily performed using the techniques of \cite{Symanzik:1972wj} and diverges logarithmically as $z \to 1$, in contrast with our $\Psi_\a(z)$.}
\beq
\boxed{
\label{eq:Psiprop}
\Psi_\a(z) = \frac{1}{2} \left[ Q(\a) k_{\half+\a}(z)  + Q(-\a) k_{\half - \a}(z)  \right],
\quad
Q(\a) = \frac{2\Gamma(-2\a)}{\Gamma^2(\thalf - \a)}\,.
}
\eeq
In what follows, it will be useful to rewrite $\Psi_\a(z)$ as
\beq
\Psi_\a(z) = {}_2F_1\!\left({{\thalf+\a,\thalf-\a}~\atop~{1}};\frac{z-1}{z} \right)
\eeq
using a hypergeometric identity. In particular, this makes it manifest that $\Psi_\a(1) = 1$. However, we have not yet inspected the asymptotics near $z=0$. Assuming that $\a$ is real, we find that $\Psi_\a(z) \limu{z \to 0}  z^{1/2-|\a|}$, which means that the functions $\Psi_\a$ have infinite norm. The only way to avoid this problem is to assume that $\a$ is imaginary. In that case, we find that $\Psi_\a$ has the following asymptotics:
\beq
\Psi_\a(z) \; \limu{z \to 0} \; |Q(\a)| \, \sqrt{z} \cos\left(  \text{Im}(\a)  \ln z + \text{const.}  \right) \qquad [ \a \in i \mbb{R}]
\eeq
implying that $\Psi_\a$ is rapidly oscillating near $z=0$. Notice that even for imaginary $\a$, the function $\Psi_\a(z)$ is real-valued, since it is symmetric under $\a \to -\a$. A plot of two different functions $\Psi_\a(z)$ is shown in Fig.~\ref{fig:plotfuncs}.

Since $\Psi_\a(z)$ oscillates near $z=0$ at a rate that depends on $\a$, it is at least plausible that ${\bexpec{ \Psi_\a,\Psi_\b } = 0}$ for $\a \neq \pm \b$, cf.\@ the Fourier transform on $\mbb{R}$. This is confirmed by an explicit computation, performed in Appendix~\ref{sec:gab}. There it is shown that the inner product $\bexpec{ \Psi_\a, \Psi_\b }$ behaves as a delta function on the imaginary axis. To be precise: if $f(\a)$ is defined on $i\mbb{R}$ and has compact support,  we have
\beq
\label{eq:testInt}
\frac{1}{2\pi i}\int_{-i\infty}^{i \infty}\!d \b  \, \bexpec {\Psi_\a, \Psi_\b } f(\b) = N(\a) \, \frac{ f(\a) + f(-\a) }{2}
\eeq
where
\beq
\label{eq:Ndef}
N(\a) = \frac{\Gamma(\a)\Gamma(-\a)}{2\pi \Gamma(\thalf+\a) \Gamma(\thalf-\a)} = \frac{|Q(\a)|^2}{2} \geq 0\,.
\eeq
Informally, Eq.~\reef{eq:testInt} shows that the functions $\Psi_\a(z)$ are plane-wave normalized, having norm $N(\a)$. The fact that the RHS of~\reef{eq:testInt} contains the sum $\half[f(\a) + f(-\a)]$ reflects that $\Psi_\a$ is even in $\a$, which carries over to the inner product $\bexpec{ \Psi_\a, \Psi_\b }$.

Summarizing, we have constructed a set of orthogonal eigenfunctions $\Psi_\a(z)$ with respect to the inner product~\reef{eq:innerProd}. Naively, we can appeal to familiar arguments of Sturm-Liouville theory to argue that these eigenfunctions form a complete set. In other words, we can decompose a given function $f \, : \, (0,1) \to \mbb{R}$ as follows:
\beq
\label{eq:sl2decomp}
\boxed{
f(z) = \frac{1}{2\pi i}\int_{-i\infty}^{i \infty} \frac{d \a}{N(\a)} \, \wh{f}(\a) \Psi_\a(z)
\quad
\Leftrightarrow
\quad
\wh{f}(\a) = \int_0^1\!\frac{d z}{z^2} \, f(z)\Psi_\a(z)\,.
}
\eeq
This formula describes how $f(z)$ is encoded by its ``spectral density'' $\wh{f}(\a)$, and vice versa. A mathematically rigorous way to obtain this identity will be described in the next section.

A sufficient condition for Eq.~\reef{eq:sl2decomp} to make sense is that $f$ be square integrable:
\beq
\bexpec{ f,f } = \int_0^1\frac{d z}{z^2} \, |f(z)|^2 < \infty\,.
\eeq
An equivalent condition (see the next section) is that
\beq
\frac{1}{2\pi i}\int_{-i\infty}^{i \infty} \frac{d \a}{N(\a)} \, |\wh{f}(\a)|^2
\eeq
is finite. In Secs.~\ref{sec:cb} and~\ref{sec:convergence} we discuss how these constraints can be loosened. 

Eq.~\reef{eq:testInt} shows that the $\Psi_\a(z)$ form a complete basis in $\a$ space. For reference, we remark that the $\Psi_\a(z)$ also obey a completeness relation in position space, namely
\beq
\label{eq:cpl2}
\frac{1}{2\pi i}\int_{-i\infty}^{i\infty} \frac{d\a}{N(\a)} \, \Psi_\a(z) \Psi_\a(w) = z^2 \, \dd(z-w)
\eeq
as can be deduced from~\reef{eq:sl2decomp}.

\subsection{Alpha space as a Jacobi transform}\label{sec:jacobi}

The alpha space transform $f(z) \mapsto \wh{f}(\a)$ is closely related to a known integral transform, known as the {Jacobi transform}. We will briefly describe this transform in the rest of this section, pointing to Refs.~\cite{flensted1973convolution,flensted1979jacobi,TK1} as a  point of entry in the mathematics literature. The Jacobi transform is an integral transform that makes use of the Jacobi functions:
\beq
\vartheta_{\a}^{(p,q)}(x) \ldef {}_2F_1\!\left({{\thalf(1+p+q) + \a,\, \thalf(1+p+q) -\a}~\atop{p+1}};- x \right), \quad x \geq 0\,.
\eeq
The parameters $p,q \geq 0$ are fixed, whereas the label $\a \in i \mbb{R}$ is allowed to vary continuously. Notice that $\dps{\vartheta^{(p,q)}_\a(x)}$ is even in $\a$, and therefore real-valued. Consider now a complex function $f(x)$, defined for $x \geq 0$, decaying sufficiently fast as $x \to \infty$. We assign to it its Jacobi transform $\mca{J}f$ as follows:
\beq
f(x) \, \mapsto \, (\mca{J}f)(\a) \ldef \int_0^\infty\!d x\,\omega_{p,q}(x)  f(x) \vartheta^{(p,q)}_\a(x)\,,
\quad
\omega_{p,q}(x) = x^p (1+x)^q\,.
\eeq
$\omega_{p,q}(x)$ plays the role of a weight function in position space. A standard result ---  see Theorem 2.3 of Ref.~\cite{TK1} --- is that $f$ can be restored from its Jacobi transform:
\beq
\label{eq:inverseJ}
f(x) = \frac{1}{2\pi i}\int_{-i\infty}^{i\infty} \frac{d \a}{\msc{N}_{p,q}(\a)} \, (\mca{J}f)(\a) \vartheta^{(p,q)}_\a(x) + \ldots
\eeq
where
\beq
\msc{N}_{p,q}(\a) = \frac{2\Gamma^2(1+p)\Gamma(\pm 2\a)}{\Gamma\big(\thalf(1+p+q) \pm \a\big)\Gamma\big(\thalf(1+p-q) \pm \a\big)}\,,
\quad
\Gamma(x\pm y) \ldef \Gamma(x+y)\Gamma(x-y)\,.
\eeq
The dots in~\reef{eq:inverseJ} indicate that depending on the values of $p$ and $q$ a finite number of terms must be added; equivalently, the integration contour in $\a$ can be deformed to pick up poles coming from $1/\msc{N}_{p,q}(\a)$.\footnote{A sufficient condition for such terms to be absent is $p + q + 1 > 0$ and $p - q + 1 > 0$.}

Properly speaking, $\mca{J}$ furnishes a map from the Hilbert space $L^2(\mbb{R}_{+},\, \omega_{p,q}(x) dx)$ to the space of functions on $i\mbb{R}$ which are normalizable  with respect to the measure $d\a/\msc{N}_{p,q}(\a)$. This map is an isometry: given two complex functions $f,g$, the following Parseval formula holds:
\beq
\int_0^\infty\!d x\, \omega_{p,q}(x) \, \overline{f(x)}g(x)= \frac{1}{2\pi i} \int_{-i\infty}^{i\infty} \frac{d \a}{\msc{N}_{p,q}(\a)} \, \overline{(\mca{J}^{}f)(\a)} (\mca{J}^{}g)(\a)\,.
\eeq
Specializing to the case $f=g$, this shows in which sense the Jacobi transform is unitary.

It is now straightforward to see that the alpha space transform for the $\sl2$ Casimir is a special case of the Jacobi transform with $p = q = 0$, after the change of variable $x \to (1-z)/z$. The precise dictionary is given by
\beq
\Psi_\a(z) = \vartheta_\a^{(0,0)}\!\left(\frac{1-z}{z} \right),
\qquad
\int_0^\infty dx\, \omega_{0,0}(x) = \int_0^1\frac{dz}{z^2}\,,
\qquad
N(\a) = \msc{N}_{0,0}(\a)\,.
\eeq
A direct consequence is the identity
\beq
\label{eq:alphaparseval}
\bexpec{f,g} = \frac{1}{2\pi i} \int_{-i\infty}^{i \infty}\frac{d \a}{N(\a)} \, \overline{\wh{f}(\a)} \wh{g}(\a)\,.
\eeq
It would be interesting to see if further theorems concerning the Jacobi transform can be recycled to prove results about alpha space densities in CFTs.

Our discussion has been quite abstract so far and at this stage the reader may want to experiment with some explicit alpha space computations. To do so, it is useful to know that the Jacobi transform essentially maps rational functions to polynomials. A precise statement is the following. Let
\beq
P_n^{(p,q)}(x) = \frac{(p+1)_n}{n!} \, {}_2F_1\!\left(-n,n+p+q+1;p+1;\thalf(1-x) \right)
\eeq
be a Jacobi polynomial of degree $n$. Then for any $r,s \geq 0$ we have~\cite{koornwinder1985special}
\begin{multline}
  \label{eq:jnaarw}
  \int_0^\infty \!d x\, \omega_{p,q}(x) \frac{1}{(1+x)^{\half(p+q+r+s)+1}} P_n^{(p,r)}\!\left(\frac{1-x}{1+x} \right) \vartheta_\a^{(p,q)}(x) \\
  = \frac{(-1)^n}{n!} \frac{\Gamma(p+1)\Gamma\big(\thalf(r+s+1) \pm \a \big)}{\Gamma\big(\thalf(p+ q+r+s)+1+n \big)\Gamma\big(\thalf(p- q+r+s)+1+n \big)}  \\
  \times \wils_n\!\left(\a;\thalf(p+q+1),\thalf(p-q+1),\thalf(r+s+1),\thalf(r-s+1)\right)\,.
  \end{multline}
The object on the last line is a Wilson polynomial~\cite{wilson1980some,AAR,askeyScheme}:
\beq
\label{eq:wdef}
\wils_n(\a;a,b,c,d) = (a+b)_n(a+c)_n(a+d)_n \,{}_4F_3\!\left({{-n,a+\a,a-\a,n+a+b+c+d-1}~\atop{a+b,a+c,a+d}};1\right).
\eeq
Evidently $\wils_n(\a;a,b,c,d)$ is a polynomial of degree $n$ in $\a^2$, and it can be shown that $\wils_n$ depends symmetrically on its parameters $a,b,c,d$. Specializing to alpha space ($p=q=0$) whilst setting $r\to 0$, $s \to 2\rho-2$, the identity~\reef{eq:jnaarw} becomes
\beq
\int_0^1\frac{d z}{z^2} \, z^{\rho} P_n(2z-1) \Psi_\a(z) = \frac{(-1)^n}{n!} \frac{\Gamma\big(\rho-\half\pm\a\big)}{\Gamma^2(\rho+n)} \, \wils_n\!\left(\a;\thalf,\thalf,\rho-\thalf,\tfrac{3}{2}-\rho\right)
\label{eq:jaco4alpha}
\eeq
where $P_n$ is a Legendre polynomial of degree $n$. This formula can be used to find the alpha space counterpart of rather general functions in position space. As a simple example, we can set $n=0$ to find the alpha space version of the function $z \mapsto z^\rho$:
\beq
\label{eq:powerlaw}
\int_0^1\frac{d z}{z^2} \, z^{\rho} \,  \Psi_\a(z) = \frac{\Gamma\big(\rho-\half\pm\a\big)}{\Gamma^2(\rho)} \,.
\eeq
An additional example will be discussed in Sec.~\ref{sec:examples}.

\subsection{Convergence of the alpha space transform}\label{sec:conv1}

Before we turn to the application of alpha space to CFTs, let us comment on the convergence of the alpha space transform $f(z) \mapsto \wh{f}(\a)$. We have in mind a function $f(z)$ that has power-law growth at $z=0$ and $z=1$, i.e.
\beq
f(z) \limu{z \to 0} z^p \qaq f(z) \limu{z \to 1} \frac{1}{(1-z)^q}\,.
\eeq
Moreover, we assume that $f(z)$ admits an expansion in powers of $z^h$ around $z=0$, meaning that it is possible to write $f(z) = \sum_{n=1}^\infty c_n z^{h_n}$. All of these conditions are certainly satisfied when $f(z)$ describes a CFT correlation function. 

Let us first consider the case where $p > \frac{1}{2}$ and $q<1$. In that case, the integral defining its alpha space density
\beq
\label{eq:int1def}
\wh{f}(\a) \ldef \int_0^1 \frac{dz}{z^2} \, f(z) \Psi_\a(z)
\eeq
converges whenever $|\Re(\a)| < p - \th$, meaning that $\wh{f}(\a)$ is holomorphic on a finite strip. Moreover, using the alpha space transform of a single power law~\reef{eq:powerlaw}, it is possible to show that $\wh{f}(\a)$ extends to a meromorphic function on the entire complex plane, with poles at $\a = h_n - \tfrac{1}{2}+ \mbb{N}$ for $n=1,2,\ldots$ (plus mirror poles on the left half plane).

Next, consider the case $p < \frac{1}{2}$, $q < 1$. In this case, it is convenient to decompose $f(z)$ as 
\beq
f(z) = f_\text{sing}(z) + f_\text{reg}(z)
\eeq
where
\beq
f_{\text{sing}}(z) = \sum_{h_n < 1/2} c_n z^{h_n}
\qaq
f_{\text{reg}}(z) = \sum_{h_n > 1/2} c_n z^{h_n}\,.
\eeq
By construction, the regular piece $f_\text{reg}(z)$ has a well-defined alpha space transform that extends to a meromorphic function on $\mbb{C}$. We can define the density $\wh{f}_\text{sing}(\alpha)$ termwise, by analytically continuing Eq.~\reef{eq:powerlaw} to arbitrary values of $\rho$.\footnote{Such analytic continuations may require a deformation of the alpha space integration contour away from the imaginary axis. Below we explain how to deal with such cases.} Concretely, we take the alpha space transform of $f(z)$ to be
\beq
\wh{f}(\a) = \sum_{h_n < 1/2} c_n \frac{\Gamma(h_n - \th \pm \a)}{\Gamma^2(h_n)} + \wh{f}_\text{reg}(\a)\,.
\eeq
If the leading term of $f_\text{sing}(z)$ is a constant, the above argument breaks down, since $1/\Gamma^2(h)$ vanishes when $h \to 0$. This is an order-of-limits issue, which can be avoided by writing $1$ as the limit of $z^\eps$ as $\eps \to 0$. 

Finally, we consider the case $q > 1$. For simplicity we consider $p > \th$, but the case of general $p$ is straightforward to treat using the above discussion. Given that $q>1$, the integral defining $\wh{f}(\a)$ diverges for all values of $\a$. We thus regulate this integral by cutting it off at $z = 1-\eps$, writing
\beq
\wh{f}_\eps(\a) \ldef \int_0^{1-\eps} \frac{dz}{z^2} \, f(z) \Psi_\a(z)\,.
\eeq
Notice that this regulator does not affect the analytic structure of $\wh{f}(\a)$: all poles originate from the region of integration near $z=0$. Now, to isolate divergent pieces in $\eps$ we notice that $\Psi_\a(z)$ admits an expansion in powers of $(1-z)$ of the following form:
\beq
\Psi_\a(z) = \sum_{k=0}^\infty s_k(\a) (1-z)^k
\eeq
where $s_k(\a)$ is a polynomial of degree $k$ in $\a^2$. This implies that $\wh{f}_\eps(\a)$ has the following structure of divergences:\footnote{To derive this formula, we are assuming that $f(z)$ admits an expansion of the form
\[
f(z) = \frac{1}{(1-z)^q} \left[ \text{const.} + \sum_{n \geq 1} a_n(1-z)^n \right]
\]
around $z=1$. If $f(z)$ rather behaves as a more general sum of power laws
\[
f(z) = \frac{c_1}{(1-z)^{q_1}} +  \frac{c_2}{(1-z)^{q_2}} + \ldots 
\]
Eq.~\reef{eq:ct1} is modified in a straightforward fashion.
}
\beq
\label{eq:ct1}
\wh{f}_\eps(\a) = \big[\text{finite as } \eps \to 0\big]  + \; \sum_{j=0}^{\floor{q-1}} \frac{t_j(\a)}{\eps^{q-1-j}}
\eeq
where $t_j(\a)$ is a polynomial in $\a$. Consequently, we take $\wh{f}(\a)$ to be the finite piece of $\wh{f}_\eps(\a)$ obtained by subtracting the divergent terms in~\reef{eq:ct1}.

\subsection{Conformal block decomposition}\label{sec:cb}

As a first application of the alpha space formalism of the previous sections, we will show that it can be used to compute conformal block decompositions for CFT correlators. As a starting point, we have in mind a meromorphic spectral density $\wh{f}(\a)$, even in $\a$, written in the following form:
\beq
\wh{f}(\a) = \sum_n \frac{-R_n}{\a - \a_n} \; + \; (\a \to -\a) \; + \; \text{entire}.
\eeq
The minus sign in front of $R_n$ is a choice of convention. We will assume that all poles $\a_n$ lie on the positive real axis; in particular, we see that every pole has a corresponding mirror pole $-\a_n$ on the negative real axis.

Our goal is to compute the position space counterpart of $\wh{f}(\a)$:
\beq
f(z) = \int_{\mca{C}} \frac{[d\a]}{N(\a)} \, \wh{f}(\a) \Psi_\a(z)
\eeq
where $\mca{C}$ is a contour parallel to the imaginary axis. Here and in what follows we write contour integrals as
\beq
\int\![d \a] = \int_{-i \infty}^{i \infty}\frac{d \a}{2\pi i}
\eeq
to avoid notational clutter. Notice that both $\wh{f}(\a)$ and the measure $N(\a)$ are even in $\a$, which means that we can replace $\Psi_\a(z)$ by any linear combination of the conformal block $Q(\a)k_{\half+\a}(z)$ and its shadow $Q(-\a)k_{\half-\a}(z)$. Without loss of generality, let us attempt to close the contour $\mca{C}$ to the right, picking up all poles $\a_n$ on the right half plane. This means that we have to drop the shadow part $\sim Q(-\a)/N(\a) \times k_{\half-\a}(z)$, as it grows exponentially on the right half plane, whereas the conformal block part decreases as $\Re(\a) \to \infty$. Consequently, we find that the position space version of $\wh{f}(\a)$ is given by
\beq
\label{eq:tpint}
f(z) = \int_{\mca{C}}\![d \a] \, \wh{f}(\a) \frac{Q(\a)}{N(\a)}k_{\half+\a}(z) = \int_{\mca{C}}\![d \a] \, \wh{f}(\a) \frac{2}{Q(-\a)}k_{\half+\a}(z)
\eeq
using the second equality in~\reef{eq:Ndef}. 
In that case, we can rewrite $f(z)$ as
\beq
\label{eq:CBdec2}
f(z) = \sum_n \frac{2R_n}{Q(-\a_n)} \, k_{\half+\a_n}(z)\,.
\eeq
To pass from Eq.~\reef{eq:tpint} to~\reef{eq:CBdec2}, we used that $1/Q(-\a)$ is analytic on the right half plane. But the sum appearing in the RHS is precisely a CB decomposition --- cf.\@ Eq.~\reef{eq:CBdec} --- where the $n$-th term corresponds to an exchanged operator $\Oo_n$ of dimension $[\Oo_n] = 1/2 + \a_n$, having OPE coefficient
\beq
\label{eq:restoOPEcoeff}
\quad \la_{\phi \phi \Oo_n}^2 = \frac{2R_n}{Q(-\a_n)}\,.
\eeq
Since $Q(-\a) > 0$ for all $\a>0$, we conclude that $ \la_{\phi \phi \Oo_n}^2$ is positive iff $R_n$ is positive.

\begin{figure}[htb]
  \begin{center}
        \includegraphics[scale=1]{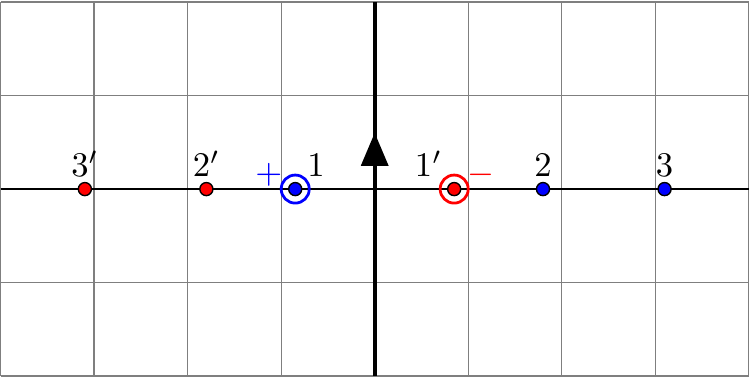}
\end{center}
\caption{\label{fig:contour} Choice of contour for a typical CFT correlator in the complex $\a$-plane. The blue dots, labeled by $\{1,2,3\}$, correspond to physical poles of $\wh{f}(\a)$, whereas their mirrors (in red, with primed labels) are unphysical. The pole 1 has $\Re(\a) < 0$, hence it corresponds to an operator of dimension $h < 1/2$. The contour runs upwards along the imaginary axis, but in this case it must circle 1 in the positive and $1'$ in the negative direction, as indicated.
  }
\end{figure}

Above, we assumed that all $\a_n$ were positive. This means that only operators of dimension $[\Oo_n] > 1/2$ appear in the CB decomposition~\reef{eq:CBdec2}. This condition can be loosened: an operator of dimension $h<1/2$ would simply correspond to a pole $\a_*$ lying on the left half plane. We must in this case deform the contour to circle $\a_*$ in the positive direction. Moreover, $\a_*$ will have a mirror pole $-\a_{*}$ on the right half plane, which must be circled in the \emph{negative} direction, such that it does not give an anomalous contribution to $f(z)$ --- see Fig.~\ref{fig:contour}. We will revisit this point in Sec.~\ref{sec:convergence}.

The cases $h=0$ (corresponding to the unit operator) and $h=1/2$ require special attention. As for $h=0$, notice that $1/Q(-\a)$ has a pole at $\a = -1/2,$ namely
\beq
\frac{1}{Q(-\a)} \; \limu{\a \to -1/2} \; -\frac{1}{\a + 1/2}\,.
\eeq
Consequently, it suffices for $\wh{f}(\a)$ to be finite at $\a = 1/2$ in order to generate a unit operator term. To be precise, if
\beq
\wh{f}(\a) \; \limu{\a \to -1/2} \; c + O[(\a+\thalf)]
\eeq
and the contour is such that it wraps around $\alpha = -1/2$ in the sense described above, then $f(z) = 2c + \text{other conformal blocks}$. A similar issue arises if $h=1/2$, because $1/Q(-\a)$ vanishes as $\a \to 0$. More precisely
\beq
\frac{1}{Q(-\a)} \; \limu{\a \to 0} \; \pi \a + O(\a^2)
\eeq
hence in order to obtain a contribution $f(z) \sim c\,  k_{1/2}(z) \sim c \sqrt{z} + \ldots$ in position space, we must have
\beq
\wh{f}(\a) = -\frac{c }{2\pi \a^2} + O(\a^{-1})\,.
\eeq

\subsubsection{Examples}
\label{sec:examples}

To develop some familiarity with the alpha space representation of correlation functions, we will compute the alpha space transform of some simple functions in $z$-space, and we use these results to compute the resulting conformal block decompositions. 

\begin{itemize}
\item Let's compute the alpha space transform of a single conformal block $k_h(z)$ with $h > 1/2$:
\beq
\label{eq:cbt}
\wh{k_h}(\a) = \int_0^1 \frac{dz}{z^2}\, k_h(z) \Psi_\a(z) = \frac{\msc{C}(h)}{\a^2 - (h-\th)^2}\,,
\quad
\msc{C}(h) = -\frac{\Gamma(2h)}{\Gamma^2(h)}\,.
\eeq
In order to derive this result, it's convenient to use the Mellin-Barnes formula
\beq
\label{eq:mbform}
{}_2F_1(a,b;c;z) = \frac{\Gamma(c)}{\Gamma(a)\Gamma(b)} \int_{-i\infty}^{i\infty} ds \, \frac{\Gamma(-s)\Gamma(a+s)\Gamma(b+s)}{\Gamma(c+s)}(-z)^s
\eeq
in order to expand both $k_h(z)$ and $\Psi_\a(z)$. Alternatively, Eq.~\reef{eq:cbt} is easy to check numerically inside the strip $|\Re(\a)| < h-\th$. 

Let us make two comments about the formula~\reef{eq:cbt}. First, although the integral in~\reef{eq:cbt} converges only in a finite strip, the RHS defines an analytic continuation to any value of $\a$. Moreover, the same formula defines an analytic continuation to values of $h < 1/2$. Second, $\wh{k_h}(\a)$ has precisely one pole on the right half plane, at $\a = h-1/2$, in accordance with our discussion from the previous section.

\item Let $f_p(z) = z^p$ with $p>1/2$. We have already encountered this function in Eq.~\reef{eq:powerlaw}, finding that in alpha space it becomes
  \beq
  \label{eq:poweralpha}
  \wh{f_p}(\a) = \frac{\Gamma(p-\thalf \pm \a)}{\Gamma^2(p)}\,.
  \eeq
  Let's use this to obtain the CB decomposition of $f_p(z)$. First, we note that $\wh{f}_p(\a)$ has poles at
  \beq
  \a_n = p - \thalf + n
  \quad \text{and} \quad
  \tilde{\a}_n = \thalf - p - n
  \qquad
  \text{with}
  \quad
  n \in \mbb{N}\,.
  \eeq
  Closing the contour to the right, we only pick up the $\a_n$ poles. The residue of the $n$-th pole is
  \beq
  R_n = - \text{Res} \; \wh{f}(\a)\big|_{\a = \a_n} = \frac{(-1)^{n}}{n!} \frac{\Gamma(2p-1+n)}{\Gamma^2(p)}
  \eeq
  and this pole corresponds to an operator of dimension $h = 1/2 + \a_n  = p + n$.
  Using the argument from the previous section, we conclude that
  \beq
  \label{eq:powerCB}
  f_p(z) = \sum_{n=0}^\infty \frac{2R_n}{Q(-\a_n)} k_{p+n}(z) = \sum_{n=0}^\infty \frac{(-1)^n}{n!} \frac{(p)_n^2}{(2p-1+n)_n}  k_{p+n}(z)\,.
  \eeq
  This confirms a known result, see for instance Eq.~(4.15) from Ref.~\cite{Gaiotto:2013nva}.
  
\item
  Let $f_{p,q}(z) = z^p(1-z)^{-q}$. It will be instructive to spend some time on the computation of the alpha space density $\wh{f}_{p,q}(\a)$. As a first step, we rewrite $\Psi_\a(z)$ using the Mellin-Barnes representation~\reef{eq:mbform}.
  This means that we can write
  \begin{align}
    \wh{f}_{p,q}(\a) &= \frac{1}{\Gamma(\thalf \pm \a)} \int\![d s] \, \frac{\Gamma(-s)\Gamma(\thalf \pm \a + s)}{\Gamma(1+s)}\int_0^1 \!\frac{d z}{z^2} \left(\frac{1-z}{z} \right)^s \frac{z^p}{(1-z)^q} \nn \\
    &= \frac{1}{\Gamma(\thalf \pm \a)\Gamma(p-q)} \int\![d s] \, \frac{\Gamma(-s)\Gamma(\thalf \pm \a + s)}{\Gamma(1+s)} \Gamma(1-q+s)\Gamma(p-1-s)
  \end{align}
  where in the first line we have interchanged the $z$ and $s$ integrals. What remains is a standard Mellin-Barnes integral, which evaluates to
  \begin{multline}
    \wh{f}_{p,q}(\a) = \frac{\Gamma(p-1)\Gamma(1-q)}{\Gamma(p-q)} \; {}_3F_2\!\left({{\thalf + \a,\, \thalf-\a,\, 1-q}~\atop~{1,\, 2-p}};1 \right)  \\
    +  \frac{\Gamma(1-p)\Gamma(p-\thalf \pm \a)}{\Gamma(\thalf \pm \a)\Gamma(p)} \;  {}_3F_2\!\left({{p-\tfrac{1}{2} + \a,\, p-\tfrac{1}{2} - \a,\, p-q}~\atop~{p,\, p}};1 \right).
  \end{multline}
which provides an analytic continuation to all $\a$, provided that ${q > p-1}$.\footnote{Interestingly, the above expression can be analytically continued to other values of $p$ and $q$ using hypergeometric identities, in particular Thm.~(2.4.4) and Corrollary~(3.3.5) from~\cite{AAR}. We can for instance write
 \[  \wh{f}_{p,q}(\a) = \frac{\Gamma(1-q)\Gamma(p-\thalf \pm \a)}{\Gamma(p)\Gamma(p-q)}\; {}_3F_2\!\left({{\thalf + \a,\, \thalf-\a,\, q}~\atop~{p,\, 1}};1 \right) = \frac{\Gamma(p - \thalf \pm \a)}{\Gamma^2(p)} \; {}_3F_2\!\left({{p-\thalf + \a,\, p-\thalf-\a,\, q}~\atop~{p,\, p}};1 \right). \]
 The ${}_3F_2(1)$ hypergeometrics in these expressions converge when $p > q$ resp.\@ $q > 1$. 
   }
 Notice that the first term above is analytic in $\a$, hence it does not contain any poles in $\a$.  However, it does influence the behaviour of $\wh{f}_{p,q}(\a)$ at large $\a$. The second term contributes two series of poles, at $\pm \a = p - \thalf + \mbb{N}$. Closing the $\a$-contour to the right and computing residues, we arrive at the following conformal block decomposition:
  \beq
  \label{eq:pqCBdec1}
  f_{p,q}(z) = \sum_{n=0}^\infty \frac{(p)_n^2}{n! (2p-1+n)_n}\, {}_3F_2\!\left({{-n,\, 2p-1+n,\,p-q}~\atop~{p,\,p}};1 \right)  k_{p+n}(z)\,.
  \eeq
  This is a new result which would have been rather difficult to guess. For $p=q$, this reduces to Eq.~(4.14) from~\cite{Gaiotto:2013nva}.
  
  \end{itemize}

\subsection{Convergence and asymptotics}\label{sec:convergence}

In Sec.~\ref{sec:conv1} we discussed the convergence of the alpha space transform in a general setting. In the present section, we will specialize to CFT correlation functions, and more particularly, we will relate the large $\a$ behaviour of $\wh{f}(\a)$ to the growth of $f(z)$ as $z \to 1$. Recall that at the extreme points $z = 0$ and $z = 1$ a crossing-symmetric four-point function in a unitary CFT behaves as
\beq
\begin{split}
z &\to 0 : \Fphi(z) \to 1 + \ldots \\
z &\to 1 : \Fphi(z) \to \left(\frac{z}{1-z}\right)^{-2h_\phi} (1 + \ldots )
\end{split}
\eeq
Clearly such a function is not square integrable with respect to the inner product~\eqref{eq:innerProd}. As we will now proceed to explain, an alpha space transform can nevertheless be defined also for such functions. We will show that divergences near the two endpoints $z = 0$ and $z=1$ translate very differently into alpha space and bear resemblance to the usual IR and UV divergences in Fourier space.

Let us first focus on $z \to 0$, which is the OPE limit, and suppose we try to transform a function $f(z)$ behaving like $z^p (1 + \ldots)$ for small $z$ to alpha space. For our inner product square integrability is lost as we dial $p$ to a value less than or equal to $1/2$. In alpha space this is reflected by a pair of poles crossing the real axis, as follows from the correspondence between conformal blocks of dimension $h$ and poles at $\alpha = \pm (h - 1/2)$. This forces the integration contour in the \emph{inverse} alpha transform off the imaginary axis, since the correct position-space expression is recovered only if it wraps around the poles as indicated in Fig.~\ref{fig:contour}. This is however the only modification necessary, and we conclude that $z \to 0$ singularities of power-law form can be entirely dealt with by augmenting the inverse alpha space transform \eqref{eq:sl2decomp} with a contour prescription around the poles. This prescription works without issues for any $0 < p < 1/2$; the special cases $p=0$ and $p=1/2$ were discussed above in Sec.~\ref{sec:cb}.

Now let us consider the limit $z \to 1$. For simplicity we will restrict ourselves to the (physically relevant) case of functions $f(z)$ analytic in $0 < z < 1$. First of all, since $\Psi_\alpha(1) = 1$ we find that
\beq
f(1) = \int [d \a] \frac{\wh{f}(\a)}{N(\a)}\,,
\eeq
and similarly it follows from $D\cdot \Psi_\alpha(z) = (\a^2 - 1/4) \Psi_\a (z)$ that
\beq
\label{eq:dersatone}
D^n \cdot f(1) = \int [d\a] (\a^2 - 1/4)^n \frac{\wh{f}(\a)}{N(\a)}\,,
\eeq
which holds as long as the $D^n \cdot f(z)$ remains square integrable. Supposing $f(z)$ behaves as a power law near $z = 1$, we see from
\beq
D \cdot \left( (1-z)^{\rho} (1 + \ldots ) \right) = \rho^2 (1-z)^{\rho - 1}(1 + \ldots)\,,
\eeq
that acting with the Casimir operator $D$ worsens the behavior near $z = 1$. For generic positive $\rho$ there exists an $n$ such that $D^n \cdot f(1)$ ceases to be well-defined, and therefore the integral in \eqref{eq:dersatone} should somehow suffer the same fate. Since we only modify the integrand with a polynomial factor, this can only happen if the integral stops converging. We conclude that the large alpha behavior reflects the `short-distance' behavior of $f(z)$ as $z \to 1$.\footnote{We can also offer a physical explanation. For fixed alpha the $\Psi_a(z)$ oscillate very slowly near $z = 1$ and to probe this region we need to consider very short `wavelengths', corresponding to very large values of the `momentum' $\alpha$.}

The above discussion also offers a way to make sense of power-law \emph{divergent} densities in alpha space: we just divide $\wh{f}(\a)$ by sufficiently powers of $\alpha^2 - 1/4$, perform the now-convergent integral over $\alpha$, and act just as many times with $D$ on the resulting position-space expression. This is in fact entirely analogous to the usual trick in Fourier space, where we habitually make sense of UV-divergent expressions like $p^{2 \alpha}$ with $\alpha > 0$ by replacing powers of $p^2$ with a Laplacian operator,
\beq
\int dx\, e^{i p x} p^{2 \alpha}( 1 + \ldots) \to (-\square)^n \left( \int dx \, e^{i p x} p^{2 \alpha - 2 n}( 1  + \ldots) \right)\,,
\eeq
with $n$ chosen such that the integral becomes convergent at large $p$. 

The relation between large $\alpha$ and $z$ close to 1 can be made more quantitative. Firstly, if a function $f(z)$ is infinitely differentiable at $z = 1$, then the preceding logic demonstrates that $\wh{f}(\a)/N(\a)$ must fall off faster than any power for large imaginary alpha. This is exemplified by the alpha space transform of $z^\rho$ given above, which falls off exponentially fast. Secondly, for the generic power-law behavior we find that if
\beq
f(z) = (1-z)^{-\rho} \left( 1 + O(1-z) \right) \qquad \text{ then } \qquad \wh{f}(\a) = (-\a^2)^{\rho-1} \frac{\G(1-\rho)}{\G(\rho)} \left( 1 + O(\a^{-2}) \right)\,,
\eeq
which can be found by subtracting the leading power using the alpha space transform of a known function. For example, for small enough $\rho$ one can use
\beq
\begin{split}
\int \frac{d^2 z}{z^2} \left[ z^{\rho}(1-z)^{-\rho} - z^\rho \right] \Psi_\a (z) &= \frac{\G(\rho - \frac{1}{2} \pm \a)}{\G^2(\rho)} \left[ \frac{\G(1-\rho)\G(\rho)}{\Gamma(\frac{1}{2} \pm \a)} - 1 \right] 
\\ &= (-\a^2)^{\rho-1} \frac{\G(1-\rho)}{\G(\rho)} \left( 1 + O(\a^{-2}) \right)
\end{split}
\eeq
which can be computed as a limit from the above examples.

\subsubsection{Application: OPE convergence}
We can use the preceding result to discuss the asymptotic behavior of OPE coefficients in one-dimensional CFTs, \emph{i.e.}, to provide a one-dimensional analogue of the results of \cite{Pappadopulo:2012jk,Rychkov:2015lca}. Such a result has been discussed previously in the context of the light-cone limit for higher-dimensional CFTs \cite{Fitzpatrick:2012yx,Komargodski:2012ek}. Here we offer an explanation based on the assumption of suitably nice asymptotic behavior in alpha space.

Consider once more a unitary CFT correlation function $\Fphi(z)$ with a corresponding alpha space expression $F(\a)$ which is meromorphic with simple poles. Our preceding discussion leads us to conclude that $F(\a) \sim (-\a^2)^{2h_\phi - 1}$ for large imaginary $\alpha$, since $\Fphi(z) \sim (1-z)^{-2 h_\phi}$ as $z \to 1$. We will assume that this asymptotic behavior holds for all non-real $\alpha$ and so the `subtracted' function
\beq
F^{(s)}(\a) \ldef (\a^2)^{- 2 h_{\phi} + 1 - \epsilon} \, F(\a) 
\eeq
vanishes asymptotically away from the real axis for any $\epsilon > 0$. This means we can write a dispersion relation for it: we write
\beq
F^{(s)}(\a) = \oint [d\b] \frac{F^{(s)}(\b)}{\a - \b}
\eeq
and push the contour away from the point $\a$. With the arcs of the contour at infinity vanishing, we find contributions only from the cuts created by the power-law prefactor and the real axis where $F(\a)$ has poles. The contributions from the cuts can be made manifestly finite by aligning them along the imaginary axis and keeping the contour some distance away from $\alpha = 0$. It follows that the contribution from the poles, which after picking up the residues can be written as
\beq
\sum_{n} (\a_n^2)^{- 2 h_\phi + 1 - \epsilon} R_n \left( \frac{1}{\a - \a_n} + (\alpha \leftrightarrow - \alpha)\right)
\eeq
is necessarily finite as well. In a distributional sense, then, we expect the residue series to behave as
\beq
\sum_n \delta(h - h_n) R_n \sim c(h_\phi) h^{4 h_\phi - 2}\,.
\eeq
By working out the example given previously we also find the prefactor:
\beq
c(h_\phi) = \frac{1}{\Gamma^2(2 h_\phi)}\,.
\eeq
We observe that the prefactor vanishes when $2 h_\phi$ is a negative integer which is precisely when the $z = 1$ singularity in $\Fphi(z)$ also disappears.

Finally we can use equation \eqref{eq:restoOPEcoeff} and to relate this result to the asymptotic behavior of the squared primary OPE coefficients themselves as
\beq
\label{eq:ourres}
\la_{\phi \phi \Oo_h}(h)^2 \sim  \frac{4^{1 - h} \sqrt{\pi} }{\Gamma^2(2 h_\phi)} \, h^{4 h_{\phi} - 3/2} \,.
\eeq
agreeing with the lightcone bootstrap result, see e.g.~\cite{Simmons-Duffin:2016wlq}.\footnote{Strictly speaking there is a factor 2 mismatch between~\reef{eq:ourres} and formula (3.8) in~\cite{Simmons-Duffin:2016wlq}, due to the fact that in the $d$-dimensional lightcone results only even spins are allowed to contribute.} It is interesting to see that the leading exponential falloff arises from the prefactor $Q(1/2 - h)$, and the falloff speed is independent of the external dimension.

\subsection{Alpha space for different external dimensions}
\label{sec:mixed}

So far we considered the case of a four-point function of identical operators. However, the Sturm-Liouville theory for the $\sl2$ Casimir operator applies just as well to four-point functions of different operators. In this section, we will briefly discuss this generalization.

Concretely, we have in mind a four-point function of primaries $\phi_i$ of dimension $h_i$, $i=1,\ldots,4$. Conformal symmetry restricts this correlator to have the following form:
\beq
\label{eq:mix4pt}
\expec{\phi_1(x_1)\phi_2(x_2)\phi_3(x_3)\phi_4(x_4)} = \left(\frac{|x_{24}|}{|x_{14}|} \right)^{h_{12}}\left(\frac{|x_{14}|}{|x_{13}|} \right)^{h_{34}} \frac{z^{h_{12}} \Fsch(z)}{|x_{12}|^{h_1 + h_2} |x_{34}|^{h_3 + h_4}} 
\eeq
for some function $\Fsch(z)$, using the shorthand $h_{ij} \equiv h_i - h_j$. The stripped correlator admits a conformal block decomposition of the following form:
\beq
\label{eq:mixedCB}
\Fsch(z) = \sum_\Oo \la_{\phi_1 \phi_2 \Oo} \la_{\phi_3 \phi_4 \Oo} \, k^{s}_{h_\Oo}(z)
\eeq
involving the mixed $\sl2$ conformal blocks
\beq
k^{s}_h(z) = z^{h+a} \, {}_2F_1(h +a,\, h + b;\, 2h; z)\,,
\qquad
a = - h_{12}\,, \; b = h_{34}\,.
\eeq
The sum in Eq.~\reef{eq:mixedCB} now runs over all operators that appear in both the $\phi_1 \times \phi_2$ and $\phi_3 \times \phi_4$ OPEs; the label `$s$' refers to this $s$-channel.

The blocks $k_h^s(z)$ are eigenfunctions of a mixed Casimir differential operator $D_{a,b}$:
\beq
  D_{a,b} \cdot f(z)   = w_{s}(z)^{-1} \, \frac{d}{d z} \!\left[ w_s(z)(1-z)z^2 f'(z) \right] +a(a+1) f(z)\,, \quad w_{s}(z) = \frac{(1-z)^{a+b}}{z^{2+2a}} \,,
  \eeq
  which means that $D_{a,b}$ is self-adjoint with respect to the inner product
\beq
\label{eq:mixedNorm}
\bexpec{f,g}_s = \int_0^1 \! d z\, w_{s}(z) \overline{f(z)}g(z)\,.
\eeq
Analyzing the relevant Sturm-Liouville problem leads to the following basis of eigenfunctions:\footnote{The PDE $D_{a,b}  f(z) = (\a^2-1/4) f(z)$ has a second solution, namely
  \[
  \frac{z^{2a}}{(1-z)^{a+b}} \; {}_2F_1\!\left({{\thalf-a+\a,\, \thalf-a-\a}~\atop~{1-a-b}};\frac{z-1}{z} \right)\,.
  \]
  This second solution ceases to be regular at $z=1$ when $a + b > 0$.}
\beq
  \Psi^{s}_\a(z)  =   {}_2F_1\!\left({{\thalf + a + \a,\, \thalf + a - \a}~\atop~{1+a+b}}; \frac{z-1}{z} \right) = \vartheta_\a^{(a+b,a-b)}\!\left(\frac{1-z}{z}\right). 
 \eeq
 In the second equality, we have rewritten $\Psi_\a^s(z)$ as a Jacobi function, to make contact with the integral transform introduced previously.

To connect the eigenfunctions $\Psi^{s}_\a(z)$ to the conformal blocks, we compute
\beq
\label{eq:sconnex}
\hspace{-4mm} \Psi_\a^{s}(z) = \half \left[ Q_{s}(\a) k^{s}_{\half+\a}(z) + (\a \to -\a) \right],
\quad
Q_{s}(\a) = \frac{2\Gamma(-2\a)\Gamma(1+a+b)}{\Gamma(\thalf + a-\a)\Gamma(\thalf+b-\a)}\,.
\eeq
As in the case of equal external dimensions, we can decompose any function $f(z)$ ---  normalizable with respect to~\reef{eq:mixedNorm} --- in terms of the functions $\Psi_\a^s(z)$, to wit:
\beq
\label{eq:sturmS}
f(z) = \int\!\frac{[d \a]}{N_{s}(\a)}\, \wh{f}(\a) \Psi^{s}_\a(z)
\quad \Leftrightarrow \quad
\wh{f}(\a) = \int_0^1\! d z \, w_{s}(z)  f(z) \Psi^{s}_\a(z)
\eeq
where
\beq
N_s(\a) = \frac{2\Gamma(\pm 2\a)\Gamma^2(1+a+b)}{\Gamma(\thalf + a \pm \a)\Gamma(\thalf + b \pm \a)} = \frac{|Q_{s}(\a)|^2}{2}\,.
\eeq
Some care must be taken when considering the $\a$ contour in Eq.~\reef{eq:sturmS}: when either $a,b \leq -\thalf$, the contour must be deformed in the Mellin-Barnes sense because of poles in the factor $1/N_s(\a)$. 

\subsubsection*{Cross channel}\label{sec:crosst}

Applying crossing symmetry to mixed four-point functions leads to a relation between two different four-point functions. In the case of the correlator $\expec{\phi_1 \phi_2 \phi_3 \phi_4}$, the bootstrap equation of interest is
\beq
\label{eq:mixcross}
F_{\phi_1 \phi_2 \phi_3 \phi_4}(z) = \left(\frac{z}{1-z} \right)^{2h_2}  F_{\phi_3 \phi_2 \phi_1 \phi_4}(1-z)
\eeq
where $F_{\phi_3 \phi_2 \phi_1 \phi_4}(z)$ is defined as in~\reef{eq:mix4pt} but with $\phi_1 \lra \phi_3$ and $h_1 \lra h_3$ exchanged. Such mixed crossing equations have been used intensively in computing scaling dimensions and OPE coefficients for the 3$d$ Ising and $O(N)$ models~\cite{Kos:2014bka,Kos:2016ysd}.

Like before, the correlator $F_{\phi_3 \phi_2 \phi_1 \phi_4}(z)$ appearing in the RHS of~\reef{eq:mixcross} admits a decomposition in conformal blocks and in plane-wave normalizable eigenfunctions of the conformal Casimir. However, care must be taken to use conformal blocks with dimensions $h_1 \lra h_3$ exchanged, and likewise for the eigenfunctions $\Psi_\a^s(z)$. To be completely explicit, this new conformal block decomposition reads:
\beq
\label{eq:mixedCBt}
F_{\phi_3 \phi_2 \phi_1 \phi_4}(z) = \sum_\Oo \la_{\phi_2 \phi_3 \Oo} \la_{\phi_1 \phi_4 \Oo} \, k^{t}_{h_\Oo}(z)
\eeq
with
\beq
k^{t}_h(z) = z^{h+a'} \, {}_2F_1(h +a',\, h + b';\, 2h; z)\,,
\qquad
a' = h_{23}\,, \; b' = h_{14}\,.
\eeq
Here and in what follows we use the `$t$' label for blocks and eigenfunctions in the $\phi_2 \times \phi_3 \to \phi_1 \times \phi_4$ channel.
The appropriate eigenfunctions in the $t$-channel are
\begin{align}
  \Psi_\a^t(z) \equiv \Psi_\a^s(z) \big|_{h_1 \lra h_3} &=    {}_2F_1\!\left({{\thalf + a' + \a,\, \thalf + a' - \a}~\atop~{1+a'+b'}};\frac{z-1}{z} \right)\\
  &= \vartheta_\a^{(a'+b',\, a'-b')}\!\left(\frac{1-z}{z}\right)
\end{align}
which satisfy
\beq
\label{eq:tconnex}
\hspace{-4mm} \Psi_\a^{t}(z) = \half \left[ Q_{t}(\a) k^{t}_{\half+\a}(z) + (\a \to -\a) \right],
\quad
Q_{t}(\a) = \frac{2\Gamma(-2\a)\Gamma(1+a'+b')}{\Gamma(\thalf + a'-\a)\Gamma(\thalf+b'-\a)}\,.
\eeq
Finally, the decomposition of a function $f(z)$ in terms of the functions $\Psi_\a^t$ reads
\beq
\label{eq:sturmT}
f(z) = \int\!\frac{[d \a]}{N_{t}(\a)}\, \wh{f}(\a) \Psi^{t}_\a(z)
\quad \Leftrightarrow \quad
\wh{f}(\a) = \int_0^1\! d z \, w_{t}(z)  f(z) \Psi^{t}_\a(z)
\eeq
where
\beq
w_{t}(z) = \frac{(1-z)^{a'+b'}}{z^{2+2a'}}
\quad
\text{and}
\quad
N_t(\a) = \frac{|Q_t(\a)|^2}{2}\,.
\eeq

\section{Crossing kernel}\label{sec:kernel}

So far, we have used Sturm-Liouville theory as a tool to represent conformal correlators as integrals over a set of basis functions $\Psi_\a$. In this section, we will use these integral representations to analyze crossing symmetry. In particular, we will compute the $d=1$ crossing kernel and exhibit its properties. 

\subsection{General case}
\label{sec:mixedxing}

Let us start by considering a mixed four-point function $\expec{\phi_1 \phi_2 \phi_3 \phi_4}$. For such a correlator, we can write down two inequivalent integral representations:
\label{eq:genrep}
\bsub
\begin{align}
  \expec{\phi_1 \phi_2 \phi_3 \phi_4} \; &\sim \; \Fsch(z) = \int\!\frac{[d \a]}{N_s(\a)} \, F_s(\a) \Psi_\a^s(z)\,,\\
  \expec{\phi_3 \phi_2 \phi_1 \phi_4} \; &\sim \; \Ftch(z) = \int\!\frac{[d \a]}{N_t(\a)} \, F_t(\a) \Psi_\a^t(z)\,.
\end{align}
\esub
The $\sim$ above denotes that we have omitted various unimportant scaling factors. The spectral density $F_s(\a)$ encodes information about the CB decomposition in the $s$-channel $\phi_1 \times \phi_2 \to \phi_3 \times \phi_4$, whereas $F_t(\a)$ describes the $t$-channel $\phi_1 \times \phi_4 \to \phi_2 \times \phi_3$.

\begin{figure}[htbp]
\begin{center}
\includegraphics[scale=1]{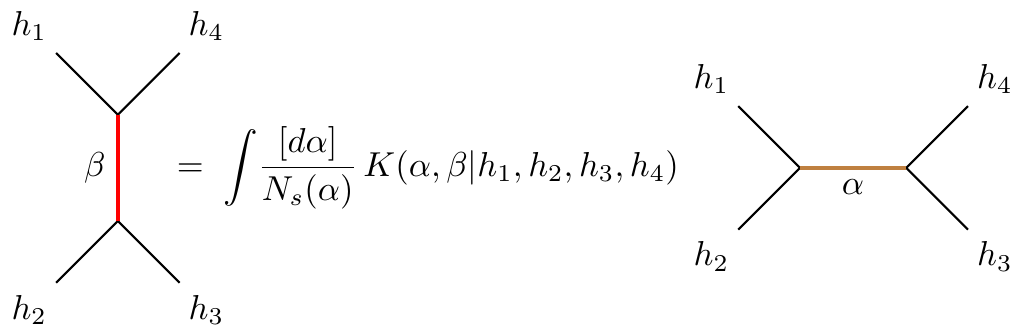}
\end{center}
\vspace{-5mm}
\caption{Graphical representation of the crossing kernel $K(\a,\b|h_1,h_2,h_3,h_4)$. \label{fig:schem}
  }
\end{figure}

The two alpha space densities $F_{s,t}(\a)$ are related --- at least implicitly --- via the crossing equation~\reef{eq:mixcross}. Plugging Eq.~\reef{eq:genrep} into that equation, we find that
\beq
\label{eq:alphaxing}
\int\!\frac{[d \a]}{N_s(\a)}\, F_s(\a) \Psi_\a^s(z) = \left(\frac{z}{1-z} \right)^{2h_2} \int\!\frac{[d\b]}{N_t(\b)} \, F_t(\b) \Psi_\b^t(1-z)\,.
\eeq
In order to find make the constraints on $F_{s,t}(\a)$ manifest, we can manipulate this alpha space bootstrap equation in various ways. For instance, it is possible to express $t$-channel eigenfunctions in terms of the $s$-channel ones:
\beq
\label{eq:kerndef}
\left(\frac{z}{1-z} \right)^{2h_2} \Psi_\b^t(1-z) = \int\!\frac{[d\a]}{N_s(\a)} \, K(\a,\b|h_1,h_2,h_3,h_4) \Psi_\a^s(z)\,.
\eeq
The distribution $K(\a,\b|h_1,h_2,h_3,h_4)$ introduced here relates eigenfunctions in the $s$- and $t$-channels, and we will refer to it as a \emph{crossing kernel}. A schematic interpretation of Eq.~\reef{eq:kerndef} is given in Fig.~\ref{fig:schem}.

 Using~\reef{eq:kerndef}, we can recast the crossing equation~\reef{eq:alphaxing} as
\beq
\label{eq:interm1}
\int\!\frac{[d\a]}{N_s(\a)} \Big[F_s(\a) - (\msf{K} \cdot F_t)(\a) \Big] \Psi_\a^s(z) = 0
\eeq
where we have introduced an integral operator $\msf{K}$ which depends on the $h_i$:
\beq
 (\msf{K}\cdot f)(\a) \ldef \int\!\frac{[d\b]}{N_t(\b)} \, K(\a,\b|h_1,h_2,h_3,h_4) f(\b)\,. 
\eeq
Recalling that the $\Psi_\a^s(z)$ form a complete basis in $z$-space, Eq.~\reef{eq:interm1} can only be satisfied if
\beq
\label{eq:xint}
F_s(\a) = (\msf{K} \cdot F_t)(\a)\,.
\eeq
The point of this identity is that it directly relates the two densities $F_{s,t}(\a)$; once we compute the kernel $K(\a,\b|h_i)$, Eq.~\reef{eq:xint} will be completely explicit. 

In the previous computation, we made an arbitrary choice by expressing $\Psi_\b^t(1-z)$ in terms of the $s$-channel functions $\Psi_\a^s(z)$. It will be useful to go in the opposite direction as well, by writing
\beq
\label{eq:kerndef2}
\left(\frac{z}{1-z} \right)^{2h_2} \Psi_\b^s(1-z) = \int\!\frac{[d\a]}{N_s(\a)} \, \wt{K}(\a,\b|h_1,h_2,h_3,h_4) \Psi_\a^t(z)
\eeq
which involves a second crossing kernel $\wt{K}(\a,\b|h_1,\ldots,h_4)$. Using the same logic as before, we arrive at an alternate alpha space crossing equation:
\beq
\label{eq:xint2}
F_t(\a) = (\wt{\msf{K}}\cdot F_s)(\a)\,,
\eeq
where
\beq
(\wt{\msf{K}}\cdot f)(\a) \ldef \int\!\frac{[d\b]}{N_s(\b)}\, \wt{K}(\a,\b|h_1,h_2,h_3,h_4) f(\b)\,.
\eeq
Bringing everything together, we have recast crossing symmetry as a system of integral equations in alpha space:
\beq
\label{eq:mixsystem}
\boxed{
  F_s(\a) = (\msf{K} \cdot F_t)(\a)\,,
  \quad
  F_t(\a) = (\wt{\msf{K}} \cdot F_s)(\a)\,.
}
\eeq

\subsection{Identical operators}

Let us briefly consider the case of the four-point function $\expec{\phi \phi \phi \phi}$ of four identical primaries. In that case, there is only one spectral density $F(\a)$ of interest, namely
\beq
\expec{\phi \phi \phi \phi} \; \sim \; \Fphi(z) = \int\!\frac{[d \a]}{N(\a)}\, F(\a) \Psi_\a(z)\,.
\eeq
Rather than a system of coupled integral equations, one now finds an eigenvalue equation for the density $F(\a)$:
\beq
\label{eq:idx}
\boxed{
  F(\a) = (\msf{K}_0 \cdot F)(\a)
}
\eeq
where the integral operator $\msf{K}_0$ is defined as
\beq
  (\msf{K}_0 \cdot f)(\a)\ldef \int\!\frac{[d\b]}{N(\b)}\, K_0(\a,\b|h_\phi) f(\b)\,,
  \quad
K_0(\a,\b|h_\phi) \ldef K(\a,\b|h_\phi,h_\phi,h_\phi,h_\phi)\,.
\eeq

\subsection{Functional properties of the crossing kernels}\label{sec:funcprop}

In what follows, we will compute the crossing kernels $K(\a,\b|h_i)$, $\wt{K}(\a,\b|h_i)$ and $K_0(\a,\b|h_i)$. Since this computation is somewhat technical, we will first derive several properties of these kernels.

Evidently, all of the kernels are even in their arguments $\a$ and $\b$. Less trivially, we see that the kernels $K$ and $\wt{K}$ are identical after exchanging the external dimensions $h_1$ and $h_3$:
\beq
\label{eq:KandKw}
\wt{K}(\a,\b|h_1,h_2,h_3,h_4) = K(\a,\b|h_3,h_2,h_1,h_4)
\eeq
as follows from Eqs.~\reef{eq:kerndef},~\reef{eq:kerndef2}.

Next, from the structure of Eq.~\reef{eq:mixsystem}, we can surmise that
\beq
\label{eq:inverses}
\msf{K} \cdot \wt{\msf{K}} = \wt{\msf{K}} \cdot \msf{K} = \id\,.
\eeq
We have derived this with input from the bootstrap, but later we will rederive Eq.~\reef{eq:inverses} formally. For the case of identical operators, Eq.~\reef{eq:inverses} becomes
\beq
\label{eq:inverses2}
\msf{K}_0^2 = \id\,.
\eeq
Notice that Eqs.~\reef{eq:inverses} and~\reef{eq:inverses2} only hold when restricted to some space of even functions, as the images of the integral operators $\msf{K}$, $\wt{\msf{K}}$ and $\msf{K}_0$ are even by construction.

Both identities~\reef{eq:inverses} and~\reef{eq:inverses2} are statements about integral operators. By acting with these operators on test functions --- say, having compact support --- we can turn them into orthogonality/completeness relations for the crossing kernels themselves. To make this concrete, let's define the distributions
\bsub
\begin{align}
\msc{D}_s(\a,\b|h_1,h_2,h_3,h_4) &\ldef  N_s(\a)^{-1} \int\!\frac{[dy]}{N_t(y)}\, K(\a,y|h_i) \wt{K}(y,\b|h_i)\,, \\
\msc{D}_t(\a,\b|h_1,h_2,h_3,h_4) &\ldef  N_t(\a)^{-1} \int\!\frac{[dy]}{N_s(y)}\, \wt{K}(\a,y|h_i) K(y,\b|h_i) \\
&= \msc{D}_s(\a,\b|h_3,h_2,h_1,h_4)\,. \nn 
\end{align}
\esub
Our claim is that $\msc{D}_{s,t}(\a,\b|h_i)$ behave as delta functions on the imaginary axis. Indeed, Eq.~\reef{eq:inverses} implies that
\beq
\label{eq:KasDirac}
\int\![d\b] \, \begin{Bmatrix} \msc{D}_s(\a,\b|h_i) \\ \msc{D}_t(\a,\b|h_i) \end{Bmatrix} f(\b) =  \frac{f(\a)+f(-\a)}{2}
\eeq
where $f(\a)$ is arbitrary. This can be thought of as the ``local'' version of~\reef{eq:inverses}. In the case of identical operators, we simply have
\beq
\label{eq:id11}
\int\![d \b] \, \msc{D}_0(\a,\b|h_\phi) f(\b) =  \frac{f(\a)+f(-\a)}{2}
\eeq
where
\beq
\label{eq:id12}
\msc{D}_0(\a,\b|h_\phi) =  N(\a)^{-1} \int\!\frac{[dy]}{N(y)} \, K_0(\a,y|h_\phi)K_0(y,\b|h_\phi)\,.
\eeq
Eqs.~\reef{eq:id11} and~\reef{eq:id12} can be obtained as a limiting case of~\reef{eq:KasDirac}. Interestingly, Eqs.~\reef{eq:KasDirac} and~\reef{eq:id11} imply that the distributions $\msc{D}_{s,t}(\a,\b|h_i)$ and $\msc{D}_0(\a,\b|h_\phi)$ are identical and independent of external dimensions $h_i$ resp.\@ $h_\phi$. As with the Fourier transform, the above identities mean that well-behaved functions $f(\a)$ can be decomposed in terms of the ``basis functions'' $K$, $\wt{K}$ and $K_0$, with computable coefficients.

\subsection{Computation of the crossing kernel}\label{sec:compu}

Let us now turn to the computation of the crossing kernel $K(\a,\b|h_i)$. To do so, we can use the alpha space technology from Sec.~\ref{sec:SL2SL} to write down a position-space integral representation for $K$, namely
\beq
\label{eq:kintrep}
K(\a,\b|h_1,h_2,h_3,h_4) = \int_0^1\!\rd z\, w_s(z)\, \left(\frac{z}{1-z} \right)^{2h_2} \Psi_\a^s(z) \Psi_\b^t(1-z)\,.
\eeq
It will be convenient to employ standard Mellin representations for the functions $\dps{\Psi_{\a}^{s,t}(z)}$:
\bsub
\begin{align}
  \Psi_\a^s(z) &= \frac{\Gamma(1+a+b)}{\Gamma\big(\thalf + a \pm \a\big)} \int\![ds] \, \frac{\Gamma(-s)\Gamma\big(\thalf+a+s\pm \a \big)}{\Gamma(1+a+b+s)} \left(\frac{1-z}{z} \right)^s \,,\\
  \Psi_\b^t(1-z) &= \frac{\Gamma(1+a'+b')}{\Gamma\big(\thalf + a'  \pm \b\big)} \int\![dt] \, \frac{\Gamma(-t)\Gamma\big(\thalf+a'+t\pm \b \big)}{\Gamma(1+a'+b'+t)} \left(\frac{z}{1-z} \right)^t \,.
\end{align}
\esub
Plugging these into \reef{eq:kintrep}, one obtains an integral representation of the form
\beq
K(\a,\b|h_i) = \int_0^1\!dz \int\![ds] \int\![dt] \ldots \,.
\eeq
Exchanging the order of the integrals, the $z$-integral yields a beta function, whereas the resulting $t$-intergral can be performed using the second Barnes lemma. What remains is the following Mellin representation:\footnote{A different-looking representation can be found by doing the $s$-integral first.}
\begin{multline}
  K(\a,\b|h_1,h_2,h_3,h_4) = \frac{\Gamma(1+a+b)\Gamma(1+a'+b')}{\Gamma\big( \thalf + a \pm \a\big)\Gamma\big(\thalf + b'\pm \b \big)}\\
  \times \;\int\![ds] \, \frac{\Gamma(-s)\Gamma\big(\thalf+a+s\pm\a \big)}{\Gamma(1+a+b+s)}  \frac{\Gamma(2h_1-1-s)\Gamma(\tfrac{3}{2}-h_1-h_4+s \pm \b)}{\Gamma(2-h_1+h_2-h_3-h_4 + s)}\,.
\end{multline}
This integral can be performed by closing the contour and picking up poles on the right half plane, at $s = \mbb{N}$ and $s = 2h_1 - 1 + \mbb{N}$. The result is a sum of two hypergeometric ${}_4F_3(1)$ functions, and it can be cast into a standard form by introducing the Wilson functions of Ref.~\cite{groenevelt2003wilson}:\footnote{Our conventions differ from those of~\cite{groenevelt2003wilson} as follows: $
      \dps{\msf{W}_\a(\b|a,b,c,d) = \phi_{i\a}(i\b;a,b,c,1-d)}$.
    }  
    \begin{multline}
      \label{eq:wilsondef}
      \hspace{-8mm} \msf{W}_\a(\b;a,b,c,d) \, =\\
      \frac{\Gamma(d-a)}{\Gamma(a+b)\Gamma(a+c)\Gamma(d\pm \b)\Gamma(\tilde{d} \pm \a)} \, {}_4F_3\!\left({{a +\b,\, a-\b,\, \tilde{a} + \a,\, \tilde{a}-\a}~\atop~{a+b,\, a+c,\, 1+a-d}};1 \right)    + (a \lra d)
\end{multline}
    writing $\tilde{a} = \thalf(a+b+c-d)$  and $\tilde{d} = \thalf(-a+b+c+d)$. It is useful to know that $\msf{W}_\a(\b;a,b,c,d)$ is even in its arguments $\a$ and $\b$, and it depends symmetrically on its parameters $\{a,b,c,d\}$. A closed-form expression for the crossing kernel is then given by
    \begin{multline}
      \label{eq:Kformula}
  K(\a,\b|h_1,h_2,h_3,h_4) =  \Gamma(1-h_{12} + h_{34})\Gamma(1+h_{14} + h_{23} )\\
  \times \; \Gamma(h_1 + h_2 - \thalf \pm \a) \Gamma(\tfrac{3}{2} - h_1 -h_4 \pm \b) \msf{W}_\a(\b;\msc{P})
\end{multline}
with parameters $\msc{P} = \msc{P}(h_1,h_2,h_3,h_4)$ specified by
\beq
\msc{P} = \left\{\half + h_{14}, \, \half + h_{23}, \, h_2 + h_3 - \half, \, \frac{3}{2} - h_1 - h_4\right\} \,.
\eeq
The kernel $\wt{K}(\a,\b|h_1,h_2,h_3,h_4)$ admits an expression similar to~\reef{eq:Kformula}, the only difference being that $h_1 \lra h_3$ are swapped. For completeness, we print the formula for the identical-operator kernel $K_0(\a,\b | h_\phi)$ here as well:
\begin{multline}
  \label{eq:podef}
  K_0(\a,\b|h_\phi) = \Gamma(2h_\phi - \thalf \pm \a)\Gamma(\tfrac{3}{2} - 2h_\phi \pm \b) \msf{W}_\a(\b;\msc{P}_0)\,,\\
  \msc{P}_0 = \left\{\half,\, \half,\, 2h_\phi - \half,\, \frac{3}{2} - 2h_\phi \right\}\,.
\end{multline}

\subsection{$\msf{K}$ and $\wt{\msf{K}}$ as intertwiners}\label{sec:intert}

Having computed the crossing kernels $K$ and $\wt{K}$, let us now revisit the alpha space crossing equation~\reef{eq:mixsystem}. Informally, it encodes that $\msf{K}$ maps a ``$t$-channel'' alpha space density to an ``$s$-channel'' one, and vice versa for $\wt{\msf{K}}$. In this section we will formalize this idea, making precise in which sense $\msf{K}$ and $\wt{\msf{K}}$ intertwine between two different Hilbert spaces. 

First, let's introduce a Hilbert space $\mca{H}_s = \mca{H}_s(h_1,h_2,h_3,h_4)$ for $s$-channel functions, consisting of all functions $f(\a)$ that are even in $\a$ and $L^2$ with respect to the following inner product:
\begin{multline}
  \label{eq:sHilb}
\big(f,g\big)_s \ldef \int\!\frac{[d\a]}{\mca{M}_s(\a;h_1,h_2,h_3,h_4)} \,\overline{f(\a)}g(\a)\,,\\
\mca{M}_s(\a;h_1,h_2,h_3,h_4) = \frac{2\Gamma^2(1-h_{12}+h_{34})\Gamma(\pm 2\a)\Gamma(h_1 + h_2 - \thalf \pm \a)}{\Gamma(\thalf - h_{12} \pm \a) \Gamma(\thalf + h_{34} \pm \a)\Gamma(\tfrac{3}{2}-h_3 - h_4  \pm \a)}\,.
\end{multline}
We have introduced an $\a$-independent factor in the measure $\mca{M}_s(\a;h_i)$ to simplify some formulas later on. The integration contour in~\reef{eq:sHilb} is to be understood in the Mellin-Barnes sense, which means that it may be deformed depending on the values of the $h_i$. Likewise, we introduce a $t$-channel Hilbert space $\mca{H}_t(h_1,h_2,h_3,h_4)$ of even functions that are square-integrable with respect to
\beq
\label{eq:tHilb}
\big(f,g\big)_t \ldef \int\!\frac{[d\a]}{\mca{M}_t(\a;h_1,h_2,h_3,h_4)} \,\overline{f(\a)}g(\a)\,,
\quad
 \mca{M}_t(\a;h_1,h_2,h_3,h_4) =  \mca{M}_s(\a;h_3,h_2,h_1,h_4)\,.
\eeq
We now claim that the following holds:

{\bf Theorem 1.1}: \emph{
   $\msf{K}$ is a unitary map $\mca{H}_t \to \mca{H}_s$, and $\wt{\msf{K}} \; : \; \mca{H}_s \to \mca{H}_t$ is its inverse.
}

\noindent 
Unitarity here means that $\msf{K}$ and $\wt{\msf{K}}$ preserve the inner products defined in Eqs.~\reef{eq:sHilb} and~\reef{eq:tHilb}, namely
\beq
\big(f,g\big)_t = \big(\msf{K}\cdot f,\msf{K}\cdot g)_s
\quad
\text{and}
\quad
\big(f,g\big)_s = \big(\wt{\msf{K}}\cdot f,\wt{\msf{K}}\cdot g)_t\,.
\eeq
The proof of this result follows from the properties of the {\it Wilson transform}, introduced in Ref.~\cite{groenevelt2003wilson}. This integral transform uses the Wilson functions $\msf{W}_\a(\b;a,b,c,d)$ as a basis. The above result can straightforwardly be deduced from Theorem 4.12 of Ref.~\cite{groenevelt2003wilson}. Consequently, we will not provide many details. However, it will be instructive to provide a sketch of a (constructive) proof. First, one establishes that $\mca{H}_s$ is spanned by the following functions:
\beq
\label{eq:xisdef}
\xi_n^{s}(\a|h_i) = \Gamma(1-h_{12} + h_{34})\Gamma(h_1 + h_2 - \thalf \pm \a) \, \wils_n(\a;\tilde{\msc{P}})\,,
\quad
n \in \mbb{N}.
\eeq
The Wilson polynomials $\wils_n$ were defined in Eq.~\reef{eq:wdef}, and the set of parameters $\tilde{\msc{P}}$ is given by
\beq
\tilde{\msc{P}}(h_1,h_2,h_3,h_4) = \left\{\half - h_{12}, \, \half + h_{34}, \, h_1 + h_2 - \half, \, \frac{3}{2} - h_3 - h_4\right\} = \msc{P}(h_3,h_2,h_1,h_4) \,.
\eeq
Likewise, $\mca{H}_t$ is spanned by the functions
\beq
\label{eq:xitdef}
  \xi_n^{t}(\a|h_i) = \Gamma(1+h_{14} +h_{23}) \Gamma(h_2 + h_3 - \thalf \pm \a) \, \mfr{p}_n(\a;\msc{P})\,.
\eeq
By linearity, it suffices to establish that $\msf{K}$ and $\wt{\msf{K}}$ act appropriately on these basis functions. To establish this, one proves first that
\beq
\label{eq:dualIP}
\big(\xi_m^{s},\xi_n^{s}\big)_s = \big(\xi_m^{t},\xi_n^{t}\big)_t \propto \dd_{mn}
\eeq
as well as
\beq
\label{eq:eigX}
(\msf{K} \cdot \xi_n^{t})(\a) = (-1)^n \, \xi_n^s(\a)\,,
\quad
(\wt{\msf{K}} \cdot \xi_n^{s})(\a) = (-1)^n \, \xi_n^t(\a)\,.
\eeq
Eq.~\reef{eq:dualIP} is a property of the Wilson polynomials $\mfr{p}_n$~\cite{AAR}, and Eq.~\reef{eq:eigX} is a consequence of Theorem 6.7\@ of~\cite{groenevelt2003wilson}.

A similar result holds for the case of identical operators. There one defines a Hilbert space $\mca{H}_0 = \mca{H}_0(h_\phi)$ of even functions that are finite with respect to
\beq
\big(f,g\big)_0 \ldef \int\!\frac{[d\a]}{\mca{M}_0(\a;h_\phi)}\,\overline{f(\a)}g(\a)\,,
\quad
\mca{M}_0(\a;h_\phi) = \mca{M}_s(\a;h_\phi,h_\phi,h_\phi,h_\phi)\,.
\eeq
Then the counterpart of the above theorem reads:

 {\bf Theorem 1.2}: \emph{ 
   $\msf{K}_0$ is a unitary map $\mca{H}_0 \to \mca{H}_0$ obeying $\msf{K}_0^2 = \mrm{id}$.
 }
 
\noindent Here unitarity means that
\beq
\big(f,g\big)_0 = \big(\msf{K}_0\cdot f,\msf{K}_0\cdot g\big)_0\,.
\eeq
The proof goes along the same lines as the general case discussed before. A basis for $\mca{H}_0$ is now spanned by the functions
\beq
\label{eq:xi0def}
 \xi^0_n(\a|h_\phi) = \Gamma(2h_\phi - \thalf \pm \a) \, \wils_n(\a;\msc{P}_0)
 \eeq
 where $\msc{P}_0$ was defined in~\reef{eq:podef}.  The operator $\msf{K}_0$ maps the $\xi_n^0$ to themselves, up to a sign $(-1)^n$:
 \beq
 (\msf{K}_0 \cdot \xi_n^0)(\a) = (-1)^n \, \xi_n^0(\a)\,.
 \eeq
Of course, the only permissible eigenvalues that could have appeared were $\pm 1$, given that $\msf{K}_0^2 = \id$.

\subsection{Analytic structure of the crossing kernel}\label{sec:analy}

Since we have rephrased bootstrap equations as integral equations in alpha space, it will be instructive to analyze the analytic structure of the crossing kernel $K(\a,\b|h_1,h_2,h_3,h_4)$. Let's first fix $\b$ and investigate the properties of $K$ as a function of $\a$, using Eq.~\reef{eq:Kformula}. Since the Wilson functions $\msf{W}_\a(\b;a,b,c,d)$ are analytic in $\a$ and $\b$, the only poles in $\a$ are due to the factor $\Gamma(h_1 + h_2 - \thalf \pm \a)$. Consequently $K(\a,\b|h_i)$ is a meromorphic function, with its only poles on the right half plane at $\a = h_1 + h_2 - \thalf + \mbb{N}$. The relevant residues are polynomials of degree $n$ in $\b^2$, namely
\begin{align}
  \label{eq:Kresgen}
  \mca{R}_n(\b;h_1,h_2,h_3,h_4) &\ldef - \text{Res}\, K(\a,\b|h_1,h_2,h_3,h_4) \big|_{\a = h_1 + h_2 - 1/2 + n} \\
  &= \frac{\Gamma(1-h_{12}+h_{34})}{n!(1+h_{14}+h_{23})_n}\frac{\Gamma(2h_1 + 2h_2 - 1+n)}{\Gamma(2h_2+n)\Gamma(h_1 + h_2 + h_{34} +n)} \nn \\
  & \hspace{20mm} \times \; \wils_n\big(\b;\thalf+h_{14},\, \thalf+h_{23},\, h_1 + h_4 - \thalf,\, h_2 + h_3 - \thalf\big)\,. \nn
\end{align}
Next, remark that for generic values of $\a$, $K(\a,\b|h_i)$ is a rather complicated function of $\b$. Upon closer inspection it appears that at certain values $\a_*$ the kernel $K(\a_*,\b|h_i)$ becomes polynomial in $\b$, up to a number of gamma functions. The relevant values $\a = \a_*$ are organized in three families:
\beq
\a_n^{\mrm{I}} = \frac{3}{2} - h_3 - h_4 + n,
\quad
\a_n^{\mrm{II}} = \half - h_{12} + n,
\quad
\a_n^{\mrm{III}} = \half + h_{34} + n,
\quad
n \in \mbb{N}\,.
\eeq
For the first family, we find for instance
\bsub
\label{eq:trunc}
\begin{align}
 K(\a_n^{\mrm{I}},\b) &= k_n^{\mrm{I}}\, \frac{\Gamma(\tfrac{3}{2} - h_1 - h_4 \pm \b)}{\Gamma(h_2 + h_3 - \thalf \pm \b)} \, \wils_n\big(\b;\thalf + h_{14},\, \thalf+h_{23},\, \tfrac{3}{2} - h_1 - h_4,\, \tfrac{3}{2}-h_2 - h_3 \big)
\intertext{where $k_n^{\mrm{I}}$ is a constant that does not depend on $\b$. For the second and third families, we find}
K(\a_n^{\mrm{II}},\b) &= k_n^{\mrm{II}}\, \frac{\Gamma(\tfrac{3}{2} - h_1 - h_4 \pm \b)}{\Gamma(\thalf + h_{14} \pm \b)} \, \wils_n\big(\b;\thalf - h_{14},\, \thalf+h_{23},\, h_2 + h_3 - \thalf,\, \tfrac{3}{2}-h_1 - h_4 \big),\\
  K(\a_n^{\mrm{III}},\b) &= k_n^{\mrm{III}}\, \frac{\Gamma(\tfrac{3}{2} - h_1 - h_4 \pm \b)}{\Gamma(\thalf + h_{23} \pm \b)} \, \wils_n\big(\b;\thalf + h_{14},\, \thalf-h_{23},\, h_2 + h_3 - \thalf,\, \tfrac{3}{2}-h_1 - h_4 \big).
\end{align}
\esub

We can also consider the analytic structure of $K(\a,\b|h_i)$ as a function of $\b$ for fixed $\a$. This is a simple exercise, given the relation~\reef{eq:KandKw}. We therefore refrain from printing explicit formulas.

\subsection{Symmetries of the crossing kernel}\label{sec:sym}

The crossing kernel obeys various identities which we will exhibit here. Since none of these results are used in the rest of this paper, this section can be skipped on a first reading.

It will be convenient to strip off the gamma functions in Eq.~\reef{eq:Kformula} and to relabel the external dimensions as $h_i \to \half + \gamma_i$. What remains is a single Wilson function, namely
\beq
\hat{K}(\a,\b|\gamma_1,\gamma_2,\gamma_3,\gamma_4) = \msf{W}_\a(\b|\thalf + \gamma_1 - \gamma_4,\, \thalf + \gamma_2 - \gamma_3,\, \thalf -\gamma_1 - \gamma_4,\, \thalf + \gamma_2 + \gamma_3)\,.
\eeq
First, we recall that $\msf{W}_\a(\b;a,b,c,d)$ depends symmetrically on its parameters $\{a,b,c,d\}$, which implies that $\hat{K}(\a,\b|\gamma_i)$ obeys
\bsub
\begin{align}
  \hat{K}(\a,\b|\gamma_1,\gamma_2,\gamma_3,\gamma_4) &= \hat{K}(\a,\b|\!-\!\gamma_1,\gamma_2,\gamma_3,\gamma_4) =  \hat{K}(\a,\b|\gamma_1,\gamma_2,-\gamma_3,\gamma_4)\\
  &= \hat{K}(\a,\b|\gamma_3,-\gamma_4,\gamma_1,-\gamma_2) \label{eq:row1}\\
  &= \hat{K}(\a,\b|\!-\!\gamma_1^\natural,-\gamma_2^\natural,-\gamma_3^\natural,-\gamma_4^\natural)\,,
  \quad
  \gamma_i^\natural =  -\gamma_i + \half \sum_{j=1}^4 \gamma_j \,.
\end{align}
\esub
A second type of symmetry can be found using the identity (see Lemma 5.3 of~\cite{groenevelt2006wilson})
\beq
\msf{W}_{\a}(\b;A+\omega,A-\omega,B+\rho,B-\rho) = \msf{W}_{\omega}(\rho;A+\a,A-\a,B+\b,B-\b)
\eeq
which descends to
\beq
\label{eq:col1}
\hat{K}(\a,\b|\gamma_1,\gamma_2,\gamma_3,\gamma_4) = \hat{K}(\gamma_1,\gamma_3|\a,\gamma_2,\b,\gamma_4) \,.
\eeq
A final relation follows from the ``duality'' property of the Wilson functions:
\beq
\label{eq:wilsdual}
  \msf{W}_\a(\b;a,b,c,d) = \msf{W}_\b(\a;\tilde{a},\tilde{b},\tilde{c},\tilde{d})\,,
  \qquad
    \begin{bmatrix} \tilde{a} \\ \tilde{b} \\ \tilde{c} \\ \tilde{d} \end{bmatrix} = \half(a+b+c+d) - \begin{bmatrix} d \\ c \\ b \\ a \end{bmatrix}
\eeq
which implies that
\beq
\label{eq:row2}
\hat{K}(\a,\b|\gamma_1,\gamma_2,\gamma_3,\gamma_4) = \hat{K}(\b,\a|\gamma_3,\gamma_2,\gamma_1,\gamma_4) \,.
\eeq

The reader may notice that the above symmetries are reminiscent of those corresponding to the $SU(2)$ $6-j$ symbol~\cite{Regge:1959ze,zhedanov1988nature,boalch2007regge,FS}. In the $SU(2)$ context, the transformations $\gamma_{1,3} \mapsto - \gamma_{1,3}$ are known as mirror symmetries and $\gamma_i \mapsto \gamma_i^\natural$ is a {Regge transformation};  Eqs.~\reef{eq:row1}~\reef{eq:col1} and~\reef{eq:row2} are related to transformations that exchange rows and columns of the $6-j$ symbol. 

A subset of the above symmetries lifts to the full crossing kernel $K(\a,\b|h_i)$:
\bsub
\label{eq:fullK}
\begin{align}
  K(\a,\b|h_1,h_2,h_3,h_4) &=K(\a,\b|h_3^\natural,h_4^\natural,h_1^\natural,h_2^\natural)\,,\quad h_i^\natural = -h _i + \half \sum_{j=1}^4 h_j\,, \label{eq:k1r}\\
  &=K(\b,\a|1-h_3^\natural,1-h_2^\natural,1-h_1^\natural,1-h_4^\natural)\,, \\
  &=K(\b,\a|1-h_1,1-h_4,1-h_3,1-h_2)\,.\label{eq:last}
\end{align}
\esub
Any two of these identities imply the third one. In conclusion, it appears that the automorphism group of the $K(\a,\b|h_i)$ is isomorphic to the Klein four-group. In passing, we note that Eq.~\reef{eq:fullK} can also be derived by inspecting the integral representation~\reef{eq:kerndef}.

\subsubsection*{Limit cases}

For bootstrap applications, one is often interested in four-point functions where some of the operators are identical. In that case, the discussion of the symmetries of the crossing kernel simplifies drastically. For a mixed four-point function of the form $\expec{\eps \sigma \sigma \eps}$, there are two relevant crossing kernels:
\beq
K_{\text{m},1}(\a,\b|h_\sigma,h_\eps) \ldef K(\a,\b|h_{\eps},h_{\sigma},h_{\sigma},h_\eps)\,,
\quad
K_{\text{m},2}(\a,\b|h_\sigma,h_\eps) \ldef K(\a,\b|h_{\sigma},h_{\sigma},h_{\eps},h_{\eps})\,.
\eeq
In this case, the content of Eq.~\reef{eq:fullK} reduces to
\beq
K_{\text{m},1}(\a,\b|h_{\sigma},h_{\eps}) = K_{\text{m},2}(\b,\a|1-h_{\eps},1-h_{\sigma})\,.
\eeq
Finally, when all external dimensions are identical, the relevant kernel is $K_0(\a,\b|h_\phi)$, which obeys
\beq
K_0(\a,\b|h_\phi) = K_0(\b,\a|1-h_\phi)\,.
\eeq

\section{Applications to the conformal bootstrap}\label{sec:appli}

In Section~\ref{sec:kernel}, we reformulated crossing symmetry in the form of integral equations in alpha space, making use of the crossing kernel $K(\a,\b|h_i)$. For definiteness, let us consider the identical-operator alpha space equation Eq.~\reef{eq:idx}:
\beq
\label{eq:eq}
F(\a) = \int\!\frac{[\rd \b]}{N(\b)} \, K_0(\a,\b|h_\phi) F(\b)\,.
\eeq
In the bootstrap context, we can ask whether Eq.~\reef{eq:eq} (combined with unitarity) can be used to find useful constraints on $F(\a)$. In this section we will sketch some ideas in this direction, making use of the properties of the crossing kernel as discussed in Sec.~\ref{sec:kernel}.

\subsection{(Dis)proving a false theorem}\label{sec:false}

We will start by outlining an simple idea for analyzing the alpha space crossing equation~\reef{eq:eq}. One can think of the RHS of~\reef{eq:eq} as a function of $\a$
\beq
\label{eq:afunc}
\alpha \, \mapsto \, \int\!\frac{[d \b]}{N(\b)}\, K_0(\a,\b|h_\phi) F(\b)
\eeq
and require that~\reef{eq:afunc} has exactly the same analytic structure as $F(\a)$, appearing on the LHS of~\reef{eq:eq}. Taken at face value, this should lead to constraints of the poles and residues of $F(\a)$, which correspond to CFT data.

The function~\reef{eq:afunc} only depends on $\a$ through the crossing kernel $K_0(\a,\b|h_\phi)$. Using the results of Sec.~\ref{sec:analy}, we see that the identical-operator kernel $K_0(\a,\b|h_\phi)$ has poles at ${\a_n = 2h_\phi - \half + n}$, $n \in \mbb{N}$, with residues
\beq
\msc{R}_n(\b|h_\phi) \ldef \mca{R}_n(\b|h_\phi,h_\phi,h_\phi,h_\phi) = \frac{\Gamma(4h_\phi - 1 + n)}{n!^2 \Gamma^2(2h_\phi + n)} \, \wils_n\!\left(\b;\thalf,\thalf,2h_\phi - \thalf,2h_\phi - \thalf\right).
\eeq
Plugging this result into~\reef{eq:eq}, we naively conclude that $F(\a)$ can only have poles at $\a = \a_n$, with their residues constrained as follows:
\beq
\label{eq:wrongthm}
- \text{Res}\; F(\a) \big|_{\a = \a_n} \stackrel{?}{=} \int\!\frac{[d \b]}{N(\b)}\, \msc{R}_n(\b|h_\phi) F(\b)\,.
\eeq

Obviously, this conclusion is wrong: it says that any solution to crossing consists of a single tower of exchanged operators with dimensions $2h_\phi + \mbb{N}$. Although solutions of this form exist (e.g.\@ in mean field theory), any interacting CFT correlator furnishes a counterexample to~\reef{eq:wrongthm}. From a mathematical point of view, we have arrived at~\reef{eq:wrongthm} using a doubtful manipulation:
\beq
\label{eq:illegal}
\text{Res} \left[ \int\!\frac{[d \b]}{N(\b)} \, K_0(\a,\b|h_\phi) F(\b) \right]_{\a = \a_n} \stackrel{?}{=}  \int\!\frac{[d \b]}{N(\b)} \, \left[ \text{Res} \; K_0(\a,\b|h_\phi) \right]_{\a = \a_n} F(\b)\,.
\eeq
This fails to hold at general $\a$, as the function~\reef{eq:afunc} is defined for real $\a$ only by analytic continuation. It would be interesting to see if this wrong argument can be refined to give useful bootstrap constraints, likely by deforming the contour in Eq.~\reef{eq:eq}, as discussed in Sec.~\ref{sec:cb}.

\subsection{Split kernel}\label{sec:split}

A second idea is to close the $\b$ contour in Eq.~\reef{eq:eq} to the right, picking up poles in $\beta$.  Since the integrand appearing in the RHS of~\reef{eq:eq} equals
\beq
\frac{K_0(\a,\b|h_\phi) F(\b)}{N(\b)}
\eeq
poles in $\b$ can come from three different factors. As mentioned, the poles in $F(\beta)$ --- and their residues --- are unknown, but of physical interest. Next, $1/N(\beta)$ has poles at $\beta = 1/2 + \mbb{N}$, and $K_0(\a,\b|h_\phi)$ has poles at $\beta = 3/2 - 2h_\phi + \mbb{N}$.\footnote{Note that the poles of $K_0(\a,\b|h_\phi)$ in $\beta$ are related to the poles in $\a$ through Eq.~\reef{eq:KandKw}. In particular, the $\b$ residues are Wilson polynomials in $\a$.} Closing the contour means that we have to keep track of all of these different poles.

We propose to modify Eq.~\reef{eq:eq} in a straightforward way, bypassing this bookkeeping exercise. The key point is that both $N(\beta)$ and $F(\beta)$ are even in $\beta$; in the definition~\reef{eq:kerndef} of the crossing kernel, it is therefore possible to replace $\Psi^t_\b(1-z)$ by $Q_t(\b) k^t_{\half+\b}(1-z)$, where $Q_t$ and $k_\a^t(z)$ were defined in Sec.~\ref{sec:crosst}. Concretely, we recast the crossing equation as
\begin{multline}
  \label{eq:spliteq}
  F(\a) = \int\![d \b] \, \Ksplit(\a,\b|h_\phi,h_\phi,h_\phi,h_\phi) F(\b)\,,\\
  \Ksplit(\a,\b|h_1,h_2,h_3,h_4) \ldef \frac{Q_t(\b)}{N_t(\b)} \int_0^1\rd z\, w_s(z) \left(\frac{z}{1-z} \right)^{2h_2} \Psi_\a^s(z) k_{\half +\b}^t(1-z)\,.
\end{multline}
We will from now on consider this ``split'' kernel $K_\text{split}(\a,\b|h_i)$ with arbitrary external dimensions, although only the case $h_1 = \ldots = h_4 \equiv h_\phi$ is of interest in the analysis of Eq.~\reef{eq:eq}.

We claim that the split kernel $\Ksplit$ does not have any poles on the right half plane $\Re(\b) > 0$. That is to say, by closing the contour of~\reef{eq:spliteq} to the right, we only pick up poles coming from $F(\beta)$, as desired.

The proof of this claim follows from a direct computation. The computation is very similar to the one from Sec.~\ref{sec:compu}. The only difference is that we use a Mellin-Barnes representation for the cross-channel block $k_h^t(1-z)$, namely
\begin{multline}
  k_{\thalf + \b}^t(1-z) = \frac{\Gamma(1+2\b)}{\Gamma(\thalf + a' + \b)\Gamma(\thalf - b' + \b)}\\
  \times \;\int\![d t]\, \frac{\Gamma(-t)\Gamma(\thalf + a'+\b+t)\Gamma(\thalf - b' +\b+t)}{\Gamma(1+2\b+t)} \left(\frac{z}{1-z} \right)^{\half+\b+a'+t}\,.
\end{multline}
As an intermediate step, we rewrite $\Ksplit$ as a Mellin-Barnes integral:
\begin{multline}
  \label{eq:splitMB}
  \Ksplit(\a,\b|h_i) = \frac{\Gamma(1-h_{12}+h_{34})}{\Gamma(1+h_{14}+h_{23})} \frac{2\b}{\Gamma(\thalf - h_{12} \pm \a)}\frac{\Gamma(\thalf + h_{23} + \b)}{\Gamma(\thalf - h_{23} + \b)} \int\![ds] \, \frac{\Gamma(-s)\Gamma(\thalf - h_{12} +s \pm \a)}{\Gamma(1-h_{12}+h_{34}+s)}
  \\
  \times \; \frac{\Gamma(2h_1 - 1 -s)\Gamma(h_{12} + h_3 + h_4 -1-s)\Gamma(\tfrac{3}{2}-h_1 -h_4 +\b+s)}{\Gamma(h_1 + h_4 - \thalf + \b-s)}\,.
\end{multline}
Closing the contour to the left\footnote{Closing the contour to the \emph{right} would mean picking up poles at $s = 2h_1 - 1 + \mbb{N}$ and $s = h_{12} + h_3 + h_4 -1 + \mbb{N}$. In the case of equal external dimensions, these two series of poles collide to form a single series of double poles.
} and picking up poles at $s = -\mbb{N}$, $s = \pm \a -\thalf + h_{12} - \mbb{N}$, we obtain the following closed-form formula for $\Ksplit$:
\beq
\label{eq:K3t}
K_{\text{split}}(\a,\b|h_1,h_2,h_3,h_4) = I_1(\a,\b|h_i) + I_2(\a,\b|h_i) +  I_2(-\a,\b|h_i)
\eeq
where
\bsub
\label{eq:Ksplitexp}
\begin{align}
  I_1(\a,\b|h_i) &=  \frac{\Gamma(1-h_{12} + h_{34})}{\Gamma(1+h_{14}+h_{23})}   \frac{2\b}{S(h_2 + h_4 + \a -\b)S(h_2 + h_4 -\a-\b)}\nn \\
  &\qquad \times \; \frac{\Gamma(\thalf + h_{14}+\b)\Gamma(\thalf + h_{23}+\b)\Gamma(\tfrac{3}{2} - h_1 - h_4 + \b)}{\Gamma(\thalf - h_{12} \pm \a)\Gamma(h_2 + h_3 - \thalf -\b)} \nn \\
  &\qquad \times \; {}_4\tilde{F}_3\!\left[{{\thalf + h_{14}+\b,\, \thalf - h_{23} + \b,\, \tfrac{3}{2} -  h_1 - h_4 + \b,\, \tfrac{3}{2} - h_2 - h_3 + \b}~\atop{1+2\b,\, 2-h_2 - h_4 + \a + \b,\, 2-h_2 -h_4 -\a + \b}};1 \right], \\
  I_2(\a,\b|h_i) &= -\frac{\Gamma(1-h_{12} + h_{34})}{\Gamma(1+h_{14}+h_{23})}  \frac{2\b}{S(h_2 + h_4 + \a-\b)} \nn \\
  & \qquad \times \; \frac{\Gamma(h_1 + h_2 - \thalf + \a)\Gamma(h_3 + h_4 - \thalf + \a)}{S(2\a)\Gamma(\thalf - h_{12} -\a)\Gamma(\thalf + h_{34} -\a)} \frac{\Gamma(\thalf + h_{23}+\b)}{\Gamma(\thalf - h_{23} + \b)} \nn \\
  &\qquad \times \; {}_4\tilde{F}_3\!\left[{{\thalf - h_{12}+\a,\, \thalf - h_{34}+\a,\, h_1 + h_2 - \thalf + \a,\, h_3 + h_4 - \thalf + \a}~\atop{1+2\a,\, h_2 + h_4 +\a+\b,\, h_2 + h_4 +\a-\b}};1 \right].
\end{align}
\esub
Here we used the notation $C(x) = \cos(\pi x)/\pi$, $S(x) = \sin(\pi x)/\pi$, and the ${}_4\tilde{F}_3(1)$ are regularized hypergeometric functions. 

Above we claimed that $\Ksplit(\a,\b|h_i)$ was analytic in $\b$ on the right half plane. This is not completely manifest from the expressions in Eq.~\reef{eq:Ksplitexp}; in fact, it appears that both $I_1$ and $I_2$ have singularities at $\b = h_2 + h_4 \pm \a + \mbb{N}$. However, it can be shown (using hypergeometric identities, see e.g.~\cite{AAR}) that the residues in $I_1(\a,\b)$ and $I_2(\pm \a,\b)$ at these points exactly cancel. Equivalently, analyticity follows from a contour pinching argument applied to the Mellin-Barnes integral in Eq.~\reef{eq:splitMB}.

In passing, we claim that $\Ksplit$ has the following symmetry:
\beq
\label{eq:kss}
  K_\text{split}(\a,\b|h_1,h_2,h_3,h_4) = K_\text{split}(\a,\b|h_3^\natural,h_4^\natural,h_1^\natural,h_2^\natural)
\eeq
cf.\@ Eq.~\reef{eq:k1r} for the normal kernel.\footnote{
  We also note the existence of a rather mysterious relation between $I_1(\a,\b|h_i)$ and $I_2(\a,\b|h_i)$, namely
\[
   I_2(\a,\b|h_1,h_2,h_3,h_4) = \frac{N_s(\a)}{N_t(\b)} \frac{C(\beta - h_{23})S(h_2 + h_4 + \a + \b)}{C(\a + h_3 + h_4) S(2\b)}  \,  I_1(\b,\a|1-h_1,1-h_4,1-h_3,1-h_2)\,.
   \]
} To establish~\reef{eq:kss}, one develops an alternate Mellin-Barnes representation for $\Ksplit$, by changing the order of integration:
\begin{multline}
  \Ksplit(\a,\b|h_i) = \frac{\Gamma(1-h_{12}+h_{34})}{\Gamma(1+h_{14}+h_{23})} \frac{2\b}{\Gamma(\thalf + h_{34} \pm \a)} \frac{\Gamma(\thalf + h_{14} +\b)}{\Gamma(\thalf - h_{14} + \b)}\\
  \int\![dt]\, \frac{\Gamma(-t)\Gamma(\thalf + a' + \b + t)\Gamma(\thalf - b' + \b + t)}{\Gamma(1+2\b+t)}  \frac{\Gamma(h_1 + h_3 - 1 - \b - t \pm \a)\Gamma(\tfrac{3}{2} - h_1 - h_4 + \b + t)}{\Gamma(h_2 + h_3 - \thalf - \b - t)}\,.
  \end{multline}
Closing the contour to the right, we find a representation of $\Ksplit$ of the schematic form~\reef{eq:K3t}, with $I_{1,2}(\a,\b|h_i)$ replaced by functions $J_{1,2}(\a,\b|h_i)$ obeying
\beq
I_k(\a,\b|h_1,h_2,h_3,h_4) = J_k(\a,\b|h_3^\natural,h_4^\natural,h_1^\natural,h_2^\natural)\,,
\quad k =1,2.
\eeq
This proves Eq.~\reef{eq:kss}.

Let us finally return to Eq.~\reef{eq:spliteq}. The modified falloff of the split kernel allows one to close the contour in the right $\beta$ plane and pick up the poles, which we have just demonstrated can only come from $F(\beta)$. Therefore, up to simple numerical factor the split kernel considered as a function of $\alpha$ for a fixed $\beta$ is precisely the $s$-channel alpha space transform of a single $t$-channel conformal block. It is therefore of interest to consider the analytic properties of $\Ksplit(\a,\b|h_i)$ in $\alpha$ as well. For example, for identical external dimensions $h_i$ a contour pinching argument applied to Eq.~\reef{eq:splitMB} shows that $\Ksplit(\a,\b|h_\phi)$ has \emph{double} rather than single poles at the double-trace values $\alpha = \pm (2 h_\phi - \hf + \mbb{N})$, reflecting the logarithmic behavior of the $k^s_{\beta + 1/2}(z)$ as $z \to 1$ in position space. This most clearly demonstrates the impossibility of expressing physical conformal blocks in one channel as proper sums of blocks in the crossed channel and consequently the necessity of using a different basis of functions like our $\Psi_\alpha(z)$ to arrive at a meaningful crossing symmetry kernel.

\subsection{Using the $\xi_n$ as a basis}

It appears that a special role is played by the alpha space functions $\xi_n^s(\a|h_i)$, $\xi_n^t(\a|h_i)$ and $\xi_n^0(\a|h_\phi)$, defined in Eqs.~\reef{eq:xisdef},~\reef{eq:xitdef},~\reef{eq:xi0def}. In fact, these basis functions furnish infinitely many solutions to crossing symmetry. To make this concrete, consider the mixed-correlator bootstrap equation~\reef{eq:mixsystem}, which is automatically solved if $F_{s,t}(\a)$ are chosen as follows:
\beq
\label{eq:mixy}
F_s(\a) = \sum_{n \text{ even}} c_n \, \xi_n^s(\a|h_i)\,,
\quad
F_t(\a) = \sum_{n \text{ even}} c_n \, \xi_n^t(\a|h_i)\,.
\eeq
It is crucial that the same coefficients $c_n$ appear both in $F_s(\a)$ and $F_t(\a)$, and that only $\xi_n$ with even $n$ appear. The reason is that the $\dps{\xi_n^{s,t}(\a)}$ with odd $n$ are antisymmetric under crossing. To understand this more intuitively, it is instructive to analyze the $\xi_n$ in position space. Using Eq.~\reef{eq:jnaarw}, we find that the $z$-space versions of $\xi_n^s(\a|h_i)$ and $\xi_n^t(\a|h_i)$ are given by
\beq
\begin{Bmatrix} \xi_n^{s}(z|h_i)  \\ \xi_n^{t}(z|h_i) \end{Bmatrix} =
n! \Gamma(2h_2 + n) \Gamma(h_1  + h_3 +h_{24} + n) \, z^{2h_2} \begin{Bmatrix} P_n^{(a+b,a'+b')}(1-2z) \\ P_n^{(a'+b',a+b)}(1-2z) \end{Bmatrix}.
\label{eq:xipos}
\eeq
Given Eq.~\reef{eq:xipos}, it follows directly that
\beq
\xi_n^{s}(z|h_i)  = (-1)^n \left(\frac{z}{1-z} \right)^{2h_2}\, \xi_n^{t}(1-z|h_i)
\eeq
where we use that $\dps{P_n^{(p,q)}(-x) = (-1)^n P_n^{(q,p)}(x)}$.  Comparing to the crossing equation~\reef{eq:mixcross}, one confirms that the $\xi_n$ with even (resp.\@ odd) $n$ are symmetric (resp.\@ antisymmetric) under crossing symmetry.

Next, we will consider the CB decomposition of the functions $\xi_n$, at least schematically. Notice that $\xi_n^s(\a|h_i)$ only has poles at $\a = h_1 + h_2 - \thalf + \mbb{N}$, as well as mirror poles on the left half plane. Given our discussion in Sec.~\ref{sec:1d}, this implies that $\xi_n^s(\a|h_i)$ has a CB decomposition consisting of operators of dimensions $h_1 + h_2 + \mbb{N}$. Such a conformal block decomposition looks similar to a mean-field solution, where only double-twist primaries $[\phi_1 \phi_2]_n \sim \phi_1 \! \stackrel{\lra}{\pd^n} \!\phi_2$ contribute. Similarly, $\xi_n^t$ has a CB decomposition with a spectrum given by $h_2 + h_3 + \mbb{N}$.  

For definiteness, we will compute the CB decomposition of $\xi^0_n(\a|h_\phi)$ explicitly. The position-space version of $\xi^0_n(\a|h_\phi)$ is a limiting case of~\reef{eq:xipos}, namely
\beq
\xi^0_n(z|h_\phi) = n! \Gamma^2(2h_\phi + n) \, z^{2h_\phi} P_n(1-2z)
\eeq
where $P_n$ denotes a Legendre polynomial. As above, these functions are crossing (anti)symmetric for even (odd) $n$, as follows from 
\beq
\xi^0_n(z|h_\phi) = (-1)^n \left(\frac{z}{1-z} \right)^{2h_\phi} \xi^0_n(1-z|h_\phi)\,.
\eeq
The CB decomposition of $\xi_n^0(z|h_\phi)$ can be found using alpha space technology; in particular, its residues in alpha space are equal to Wilson polynomials evaluated at certain values of $\a$. The precise result is
\beq
\label{eq:xicb}
\xi^0_n(z|h_\phi) = \sum_{m=0}^\infty A_m^{(n)} k^{\phantom{()}}_{2h_\phi + m}(z)
\eeq
where
\beq
A_m^{(n)} = \Gamma^2(2h_\phi+n)n! \, \frac{(-1)^m}{m!}\frac{(2h_\phi)^2_m  }{(4h_\phi-1+m)_m} \, {}_4F_3\!\left({{-n,-m,n+1,4h_\phi - 1+m}~\atop{2h_\phi,2h_\phi,1}};1 \right).
\eeq
Notice that the coefficients $\dps{A_m^{(n)}}$ are sign-alternating: $\text{sgn}( \dps{A_m^{(n)}}) = (-1)^m$, provided that $h_\phi > 0$. This implies that the $\xi_n$ do not correspond to unitary solutions of crossing.

At least formally, it is possible to derive selection rules for alpha space densities using the functions $\xi_n$. We will focus on the identical-operator case for simplicity. Recall that the $\xi_n^0$ form a basis of the Hilbert space $\mca{H}_0$ introduced in Sec.~\ref{sec:intert}. This implies that if a density $F(\a) \in \mca{H}_0$ is crossing symmetric, it must obey
  \beq
  \label{eq:selrule}
  \int\!\frac{[d \a]}{\mca{M}_0(\a;h_\phi)} \; \xi_n^0(\a|h_\phi) F(\a) = 0
  \quad \text{for} \quad n = 1,3,5,\ldots 
\eeq
This selection rule manifestly holds if $F(\a)$ is of the following form:
\beq
\label{eq:evendec}
{F}(\a) = \sum_{n \text{ even}} c_n \, \xi_n^0(\a|h_\phi) 
\eeq
cf.\@ Eq.~\reef{eq:mixy}. Of course, requiring that $F(\a)$ is normalizable imposes constraints on the growth of the coefficients $c_n$ as $n \to \infty$. 

Unfortunately an alpha space density of the form~\reef{eq:evendec} cannot belong to an interacting CFT: it would have a CB decomposition with exchanged operators of dimensions $2h_\phi + \mbb{N}$ and nothing else --- in particular, requiring that $F(\a) \in \mca{H}_0$ rules out an identity operator contribution. These unphysical constraints on the spectrum of $F$ are very similar to the issue encountered in Sec.~\ref{sec:false}. We also stress that~\reef{eq:evendec} generically corresponds to a non-unitary CB decomposition, in line with our remarks below Eq.~\reef{eq:xicb}.  Imposing unitarity leads to additional constraints on the coefficients $c_n$, and in future work it would certainly be interesting to examine these in detail.

To better understand the role played by the $\xi_n$, we will briefly consider how these ideas apply to a mean-field correlator:
\beq
F_{\text{MFT}}(z) = t_1 F_1(z) + t_2 F_2(z)
\eeq
with
\beq
F_1(z) = z^{2h_\phi}
\quad
\text{and}
\quad
F_2(z) = 1 +  \left(\frac{z}{1-z} \right)^{2h_\phi}\,.
\eeq
Both pieces $F_{1,2}$ are crossing symmetric by themselves, but only their combination with $t_2 \pm t_1 \geq 0$ is unitary. This follows from the CB decompositions~\reef{eq:powerCB} and~\reef{eq:pqCBdec1}.\footnote{
  Here we are interested in the case $p=q$ of Eq.~\reef{eq:pqCBdec1}, which reads   \[ \left(\frac{z}{1-z} \right)^p = \sum_{n=0}^\infty \frac{(p)_n^2}{n! (2p-1+n)_n} \, k_{p+n}(z)\,. \]   } Separately, $F_1$ and $F_2$ contain contributions from an infinite tower of operators of dimension $2h_\phi + n$, but the contributions for odd (resp.\@ even) $n$ cancel out when $t_1 = t_2$ (resp.\@ $t_1 = -t_2$). The combinations with $t_1 = \pm t_2$ correspond to generalized free fields with bosonic (resp.\@ fermionic) statistics.

Can we decompose $F_1$ and $F_2$ {\it \`{a} la} Eq.~\reef{eq:evendec}? As for $F_1$, we see by inspection that
\beq
F_1(z) = \frac{1}{\Gamma^2(2h_\phi)} \, \xi_0^0(z|h_\phi)
\eeq
consistent with the fact that $F_1$ is crossing symmetric and non-unitary. In particular, this shows that ${F}_1(\a) \in \mca{H}$. Notice that this is only possible because $F_1(z)$ has no unit operator contribution. Since $F_2(z)$ does have a unit operator contribution, it follows that $F_2(z)$ cannot be decomposed as in Eq.~\reef{eq:evendec}.
Nevertheless, we compute
\beq
\label{eq:decLeg}
\left(\frac{z}{1-z} \right)^{2h_\phi} = \sum_{n=0}^\infty f_n \, \xi_n^0(z|h_\phi)\,,
\quad
f_n =  \frac{1}{\Gamma^2(2h_\phi)} \frac{1+2n}{n!(1-2h_\phi+n)} \frac{1}{(2h_\phi - n)_{2n}} \,.
\eeq
Strictly speaking this holds only for $h_\phi < 1/2$; for generic $h_\phi$,~\reef{eq:decLeg} makes sense only after analytic continuation. 
Notice that~\reef{eq:decLeg} contains terms with both even and odd $n$. This is consistent with the fact that $\left[z/(1-z)\right]^{2h_\phi}$ by itself has no definite crossing behaviour. Another interesting feature is that the $f_n$ are not sign-definite; in fact, $\text{sgn}(f_n) = (-1)^n$ provided that $h_\phi < 1/2$. However, we know from Eq.~\reef{eq:pqCBdec1} that $\left[z/(1-z)\right]^{2h_\phi}$ has a CB decomposition with positive coefficients. We conclude that there is a conspiracy between the coefficients $f_n$ from Eq.~\reef{eq:decLeg} and the $\dps{A_m^{(n)}}$ from~\reef{eq:xicb} that guarantees that the full CB decomposition is unitary.

The above example shows how the idea to draw selection rules from the $\xi_n$ runs into problems when naively applied to CFT correlators. Nonetheless, it may be true that a modified version of Eq.~\reef{eq:selrule} holds after carefully regulating the identity operator contribution. We leave this question for future work.

\section{Discussion}\label{sec:discussion}

This paper has outlined how Sturm-Liouville theory provides a framework to study CFTs. Inspired by classic results \cite{dobrev1977harmonic}, we discussed the decomposition of a CFT four-point correlator in terms of a new basis of functions $\Psi_\alpha(z)$ and explained how the familiar conformal block decomposition can be obtained by analytic continuation in $\alpha$. The alpha space decomposition allowed us to formulate crossing symmetry in terms of an eigenfunction problem for some integral kernels: in particular equation \eqref{eq:eq} is a mathematically precise version of the abstract idea expressed by equation \eqref{eq:wrong} in the introduction. It features an explicitly known crossing symmetry kernel $K_0(\alpha,\beta|h_\phi)$ whose properties we analyzed in some detail.

In this paper we did not touch on the profound connection between the alpha space construction and the representation theory of the conformal group. Roughly speaking the dictionary is well-known: three-point functions map to Clebsch-Gordan kernels, conformal blocks are their square --- as used in three-fold tensor products --- and the crossing symmetry kernel is equal to a $6-j$ symbol for the conformal group. Moreover, the alpha space decomposition ought to correspond to tensor product decomposition into a direct integral over the principal unitary series of representations. We can however only make all these relations precise if we have a detailed knowledge of both the groups, the representations under consideration, and the Hilbert space of functions on which they act.\footnote{In this context it is important to note that the representations are only unitary in Lorentzian signature. In that case the conformal group is actually the universal cover of $SL(2,\mbb{R})$ \cite{Luschermack}, which has a richer class of inequivalent unitary representations \cite{Pukanszky1964} (see also \cite{Mack:1975je} for a detailed discussion of the 4$d$ case).} For the case at hand the question appears to be partially solved in \cite{groenevelt2006wilson}, which showed that the Wilson functions $\msf{W}_{\a}(\b;a,b,c,d)$ indeed appear as $6-j$ symbols for representations of the $\mathfrak{sl}(2,\mbb{R})$ conformal algebra. Surprisingly this connection works provided three of the four external dimensions transform in the \emph{discrete} unitary series, in contrast with the older discussion of \cite{dobrev1977harmonic} which is based entirely on the \emph{principal} unitary series.\footnote{This is related to our basis functions being different form the usual shadow-symmetric blocks of \cite{Dolan:2000ut} which are in fact the correct squared Clebsch-Gordan coefficients for three unitary principal series.} It would be interesting to build on the results of \cite{groenevelt2006wilson} to explicitly connect all the dots between alpha space, one-dimensional unitary CFTs and representation theory. We hope to return to this problem in the near future.

It is of clear interest to generalize our analysis to $d \geq 2$ dimensions. This requires solving the Sturm-Liouville problem for the $d$-dimensional Casimir~\cite{DO2} on the square $(0,1) \times (0,1)$, or alternatively one could relate this kernel to a suitable set of $6-j$ symbols of the universal cover of $SO(d,2)$. The higher-$d$ alpha space picture will necessarily be more complicated, because both external and exchanged operators in higher-$d$ CFTs can carry a nontrivial Lorentz spin. An obvious generalization pertains to superconformal field theories in various $d$~\cite{Cordova:2016emh}. Sturm-Liouville theory should also apply beyond four-point correlators in CFTs on $\mbb{R}^d$; for instance, one can consider its application to CFTs in the presence of boundaries or defects.
 
Most of these problems are rather formal and group-theoretical in nature. In the framework of the conformal bootstrap, it is more exciting to investigate whether alpha space crossing equations can be leveraged to constrain CFT data, or --- more ambitiously ---  to solve bootstrap equations analytically.\footnote{See~\cite{Isachenkov:2016gim} for a connection between the conformal Casimir and integrability, which may be helpful in this context.}  In Sec.~\ref{sec:appli} we discussed some tentative ideas in this direction. Together with recent developments in the realm of Mellin space and the lightcone bootstrap, we are optimistic that alpha space can become part of the analytic bootstrap toolkit.

\section*{Acknowledgments}
This work originated from discussions with Leonardo Rastelli and Pedro Liendo in Stony Brook in 2011, and we would like to thank them for their valuable contributions in these early stages. Moreover, we gratefully acknowledge discussions with the participants of the `Back to the Bootstrap II' meeting in 2012 where an initial version of this work was first presented. We would like to thank Christopher Beem, Jyoti Bhattacharya, Liam Fitzpatrick, Simon Caron-Huot, Abhijit Gadde, Leszek Hadasz, Christoph Keller, Zohar Komargodski, Hugh Osborn, Slava Rychkov, Volker Schomerus, David Simmons-Duffin and Sasha Zhiboedov for more recent discussions and/or comments. This research was supported in part by Perimeter Institute for Theoretical Physics. Research at Perimeter Institute is supported by the Government of Canada through Industry Canada and by the Province of Ontario through the Ministry of Research and Innovation. This work was additionally supported by a grant from the Simons Foundation (\#488659).

\appendix

\section{Computing the inner product $\bexpec{\Psi_\a,\Psi_\b}$}
\label{sec:gab}

In this section, we will prove Eq.~\reef{eq:testInt} by computing the inner product $\bexpec{\Psi_\a,\Psi_\b}$, as defined in Eq.~\reef{eq:innerProd}. Concretely, we must perform the following integral: 
\beq
\label{eq:inttodo}
\bexpec{\Psi_\a,\Psi_\b} = \int_0^1 \frac{d z}{z^2} \, \Psi_\a(z) \Psi_\b(z)
\eeq
where we used that $\overline{\Psi_\a(z)} = \Psi_\a(z)$ for imaginary $\a$. As a first step, we write $\Psi_\a(z)$ and $\Psi_\b(z)$ using a Mellin-Barnes representation:
\beq
\Psi_\a(z) = \frac{1}{\Gamma(\thalf \pm \a)} \int\![d s]\, \frac{\Gamma(-s) \Gamma(\thalf+s\pm \a)}{\Gamma(1+s)} \left(\frac{1-z}{z} \right)^{s}\,.
\eeq
Naively, the $z$-integral~\reef{eq:inttodo} is logarithmically divergent, the divergence coming from the region near $z = 0$. To resolve this divergence, we regulate $\Psi_\b(z)$ by writing it as follows:
\beq
  \Psi_\b(z)  \to z^\eps \; {}_2F_1\!\left({{\thalf+\b,\, \thalf-\b}~\atop~{1+\eps}};-\frac{1-z}{z} \right) =  z^{\eps} \, \frac{\Gamma(1+\eps)}{\Gamma(\thalf \pm \b)} \int\![d t]\, \frac{\Gamma(-t) \Gamma(\thalf+t\pm \b)}{\Gamma(1+\eps+t)} \left(\frac{1-z}{z} \right)^{t} 
  \eeq
for $\eps > 0$. This behaves as $O(z^{1/2+\eps})$ at small $z$. Evidently, in the limit $\eps \to 0$, the above function reduces to $\Psi_\b(z)$.

At this point, the inner product Eq.~\reef{eq:inttodo} is given by triple integral, schematically
\beq
\bexpec{\Psi_\a,\Psi_\b} = \int_0^1 d z \int[d s] \int[d t] \; ( \ldots )\,.
\eeq
Since we have regulated the integrand, this integral converges and we can exchange the order of the different integrals. We do the $z$-integral first, which is a simple beta function integral. The result is
\begin{multline}
  \ldots = \frac{\Gamma(1+\eps)}{\Gamma(\thalf \pm \a)\Gamma(\thalf \pm \b)} \int\![d s] \, \frac{\Gamma(-s) \Gamma(\thalf\pm s + \a)}{\Gamma(1+s)}\\
  \times \; \int\![d t] \, \frac{\Gamma(-t) \Gamma(\thalf+t\pm \b)}{\Gamma(1+\eps+t)} \frac{\Gamma(1+s+t)\Gamma(-1-s-t+\eps)}{\Gamma(\eps)}\,.
\end{multline}
We now do the $t$-integral, using the second Barnes lemma. This yields
\beq
\bexpec{\Psi_\a,\Psi_\b} = \lim_{\eps \to 0} \; \frac{\Gamma(1+\eps)}{\Gamma(\thalf \pm \a)\Gamma(\thalf +\eps \pm \b)} \int\![d s]\;  \frac{\Gamma(-s) \Gamma(\thalf\pm s + \a)\Gamma(-\thalf -s + \eps \pm \b)}{\Gamma(-s+\eps)}\,.
 \eeq
 At this stage we can take the limit $\eps \to 0$ everywhere, except in the two factors $\Gamma(-\thalf -s + \eps \pm \b)$:
 \beq
 \ldots = \frac{1}{\Gamma(\thalf \pm \a)\Gamma(\thalf \pm \b)} \int\![d s] \, \Gamma(\thalf + s \pm \a)\Gamma(-\thalf - s + \eps \pm \b)\,.
 \eeq
 This integral can be computed using the first Barnes lemma, yielding
 \begin{multline}
 \bexpec{ \Psi_\a, \Psi_\b } = \frac{1}{\Gamma(\thalf \pm \a)\Gamma(\thalf \pm \b)} \,  \lim_{\eps \to 0} \, Z_\eps(\a,\b)\,, \\
Z_{\eps}(\a,\b) =  \frac{1}{\Gamma(2\eps)} \Gamma(\a+\b+\eps)\Gamma(\a-\b+\eps)\Gamma(-\a+\b+\eps)\Gamma(-\a-\b+\eps)\,.
 \end{multline}
 To conclude, we need to analyze the limit $\eps \to 0$ of $Z_{\eps}(\a,\b)$, which we claim is the sum of two Dirac delta functions:
 \beq
 \label{eq:toProveZ}
 \lim_{\eps \to 0} \, \int\![d \b] \, Z_\eps(\a,\b) f(\b) = \Gamma(\pm 2\a) \left[ f(\a)+f(-\a) \right]\,,
 \eeq
 where $f(\a)$ is a test function.
 Notice that Eq.~\reef{eq:toProveZ} is sufficient to establish Eq.~\reef{eq:testInt}, after remarking that
  \beq
 \frac{2\Gamma(\pm 2\a)}{\Gamma(\thalf \pm \a)^2} = N(\a)\,.
 \eeq

 The proof of~\reef{eq:toProveZ} goes as follows. We start by noticing that $\lim_{\eps \to 0} Z_\eps(\a,\b)$ vanishes, unless $\b = \pm \a \pm n$ for some integer $n$. If $n \neq 0$, the limit $\eps \to 0$ is finite, hence such points do not contribute to the integral in Eq.~\reef{eq:toProveZ}. Hence it suffices to consider the cases $\beta =  \a$ and $\beta = -\a$. For concreteness, let's consider $\beta = \a$, in which case we can approximate $Z_\eps(\a,\b)$ by
 \beq
 Z_\eps(\a,\b)  \; \limu{\b \to \a} \; \Gamma(\pm 2\a) \, \omega_\eps(\a-\b)\,,
 \quad
 \omega_\eps(\a) = \frac{\Gamma(\eps \pm \a)}{\Gamma(2\eps)}\,.
 \eeq
 It is straightforward to see that $\omega_\eps(\a)$ behaves as a delta function along the imaginary axis, i.e.\@
 \beq
 \lim_{\eps \to 0} \; \int\![d \a] \, \omega_\eps(\a) f(\a) = f(0)\,.
 \eeq
 This follows from the fact that $\omega_\eps(\a)$ is peaked around $\a = 0$ with width $\eps$ (taking $\a$ to be imaginary) together with the fact that
 \beq
 \int\![d \a] \, \omega_\eps(\a)  = \frac{1}{4^\eps} \; \to \; 1\,.
 \eeq
 The same argument holds for the region where $\b = -\a$. This allows us to conclude.

{
\bibliographystyle{utphys}
\bibliography{biblio}
}

\end{document}